\documentclass[notitlepage,aps,prl,nofootinbib,showpacs,floatfix]{revtex4-1}

\usepackage{graphicx}
\usepackage{amsmath}
\usepackage{amssymb}
\usepackage{color}
\usepackage{txfonts}
\usepackage{verbatim}
\usepackage{lipsum}
\usepackage{subfig}

\begin{document}

\title{Framework for describing perturbations to the cosmic microwave\\background from a gravitational wave burst with memory}
\author{Dustin R. Madison}
\email{dustin.madison@mail.wvu.edu}
\affiliation{Department of Physics and Astronomy, West Virginia University, P.O. Box 6315, Morgantown, West Virginia 26506, USA}
\affiliation{Center for Gravitational Waves and Cosmology, West Virginia University, Chestnut Ridge Research Building, Morgantown, West Virginia 26505, USA}

\date{\today}
\begin{abstract}
Gravitational wave bursts with memory (BWMs) can generate measurable, long-lived frequency shifts and permanent angular deflections in distant sources of light. These perturbations vary across the sky with a characteristic spatial pattern and evolve slowly over long periods of time. In this work, we develop formalism that can be used to describe how a BWM influences the spatial pattern of temperature fluctuations in the cosmic microwave background (CMB). We limit our attention to planar gravitational wave fronts---this assumption dramatically simplifies the necessary calculations. Using toy models of the CMB's primary temperature variation pattern, we demonstrate that a BWM can mix power from a spherical harmonic mode of a certain degree into modes of various other degrees with vastly different $l$ values. In other words, BWM-induced perturbations to the CMB at any angular scale depend in detail on the unperturbed character of the CMB on all angular scales. The tools developed herein will greatly facilitate future analyses of BWM-induced temperature perturbations that incorporate all of the important physics underlying the CMB.   
\end{abstract}
\maketitle

\section{I. Introduction}
In their wake, gravitational waves (GWs) leave lasting strains in space commonly called ``memory." Memory is generated over the entire past history of gravitationally radiating systems, whether they be slowly inspiralling binaries, merging galaxies, or cosmic strings. But rapid growth of memory is tied to luminous GW events like the final mergers of binary black holes. These events are called GW bursts with memory (BWMs)  \cite{bt87,c91,t92,bd92,f09,f10}. The very existence of GW memory is a consequence of the field-theoretic properties of general relativity and the asymptotic symmetries of spacetime. Observations of memory and an improved understanding of it could have bearing on the long-standing black hole information paradox and could open windows onto necessary modifications to our understanding of gravity \cite{sz16,s17,hps16,hps17,hiw17,fn17,dn16,k20}. 

Unfortunately detecting memory with well-known methods may prove difficult. Ground-based detectors will likely only be able to infer the presence of memory in a statistical sense after thousands of GW events are detected \cite{ltl+16,htl+20,bns20}. Pulsar timing arrays could certainly detect a BWM generated by the merger of supermassive black holes \cite{cj12,mcc14,whc+15,abb+15,aab+20}, but such a merger would have to involve among the most massive black holes thought to exist, occur quite close to Earth, and happen during the decadal time span of the pulsar timing array project---such events are anticipated to occur only once per few million years \cite{isb+19}. 

In a recent paper, we discussed a possible new path forward in the pursuit of GW memory that has substantial crossover with precision cosmological studies \cite{m20}. GWs produce both redshifts and angular deflections in distant sources of light (see, e.g., \cite{bf11} and references therein). In \cite{m20}, we described the pattern of deflections generated by a planar BWM and demonstrated that those deflections last indefinitely. As such, the cosmological history of BWMs will induce deflections in distant sources of light that grow in the fashion of a random walk as more and more GW events occur. The largest possible manifestation of this signal is encoded in the distribution of the oldest light in the Universe: the cosmic microwave background (CMB). When applying memory considerations to the CMB, the redshifts induced by a BWM must also be taken into account; though the redshifts are not strictly permanent in the same way the deflections are, every BWM that has occurred since recombination is still causing a small redshift over some potentially large part of the CMB. These redshifts must be accounted for along with the deflections. 

In this paper, we have begun the task of describing how a BWM, both the redshifts and deflections that it causes, influences observable features of the CMB. We continue to operate in the planar limit that we employed in \cite{m20}. The need to eventually move beyond the planar approximation is clear as the sources producing BWMs lie between us and the surface of last scattering for the CMB. But the planar approximation is a mathematical expediency, acts as a useful tool for developing intuition pertaining to this problem, and should closely mimic the results of a full spherical wave front treatment in all but a region of sky subtending the BWM source (see \cite{mcc17} for a treatment of memory-induced redshifts from spherical wave fronts). 

In Sec. II, we present the deflection and redshift effects of a BWM in a way that facilitates our further analysis. In Sec. III, we recapitulate some well-known formalism describing the spatial pattern of temperature fluctuations in the CMB and adapt it to our purposes. In Sec. IV, we describe how the effects of a BWM introduced in Sec. II influence the CMB observables discussed in Section III and explicitly derive how BWM-induced redshifts affect the projection of the CMB's temperature fluctuation pattern onto spherical harmonics; a similar treatment of the deflection effects is reserved to a series of appendixes. Finally, in Sec. V, we provide some straightforward, informative demonstrations of our formalism and discuss forthcoming applications of this framework. 

\section{II. Effects of a BWM}
The observed changes in a frequency induced by a planar GW were first described by Estabrook and Wahlquist \cite{ew75}. Their result is at the heart of all searches for GWs with pulsar timing arrays. Specific consideration of BWMs in the pulsar timing array context began with Seto and van Haasteren and Levin \cite{s09,vl10}. The redshift pattern for a ``$+$" polarized BWM of strain amplitude $h_M$ propagating in the positive $z$ direction can be expressed as
\begin{eqnarray}
\label{eq:redshift}
z(t,\theta,\phi)&=&\frac{h_M}{4}(e^{-2i\phi}+e^{2i\phi})(1-\cos{\theta})\Theta(\theta_t-\theta),
\end{eqnarray}
where $\cos{\theta_t}=\beta-1$, $\beta=ct/d$, and $d$ is the distance to the source of light being redshifted when it emitted the light. We limit our attention to $0\leq\beta\leq2$, the span of times over which BWM-induced perturbations evolve. When $\beta=0$, the redshift pattern influences the entire sky. As $\beta$ grows with time, the redshift pattern shuts off for values of $\theta>\theta_t$. When $\beta=2$, $\theta_t=0$ and the redshift perturbation will have shut off over the entire sphere. However, we note that $\beta=2$ implies that an amount of time equal to twice the light travel time to the redshifted source has elapsed since the BWM first encountered the observer. As we are talking about the CMB and surface of last scattering, we are in reality concerned with $\beta<2$ for most BWM sources. 

The redshift signal model in Eq. (\ref{eq:redshift}) is appropriate if the memory can be treated as having turned on or built up over a short timescale. In \cite{m20}, we modeled memory signatures with a more sigmoidlike function that was parametrized by a timescale over which the memory signal developed. With that, we demonstrated that the details of the development of the signal do not matter once the rising edge of the signal has surpassed the observer. In other words, the Heaviside unit step function, $\Theta$, provides an adequate functional description of the signal if we do not care to resolve the actual buildup of the memory. We use it here as it dramatically simplifies the necessary calculations.

A rigorous description of both the redshifts and astrometric deflections induced by a GW can be found in \cite{bf11} by Book and Flanagan. In \cite{m20}, we built on the work of Book and Flanagan to describe the specific pattern of astrometric deflections from a planar BWM propagating in the positive $z$ direction. A source of light that is initially in the direction $\hat{\bf n}$ with angular coordinates $\theta$ and $\phi$ will appear deflected by a small angle
\begin{eqnarray}
\label{eq:deflection}
\delta{\bf n}(t,\hat{\bf n})&=&h_M\left\{{\bf V}_\oplus(\hat{\bf n})-{\bf V}_\bigstar(\hat{\bf n})\left[\frac{\beta\Theta(\theta_t-\theta)}{(1+\cos{\theta})}+\Theta(\theta-\theta_t)\right]\right\},~{\rm where}
\end{eqnarray}
\begin{eqnarray}
\label{eq:earthterm}
{\bf V}_\oplus&=&-\frac{1}{4}\sin{\theta}\left[(e^{2i\phi}+e^{-2i\phi})\hat{\theta}+i(e^{2i\phi}-e^{-2i\phi})\hat{\phi}\right],~~{\rm and}\\
\label{eq:starterm}
{\bf V}_\bigstar&=&-\frac{1}{4}\sin{\theta}\left[(1+\cos{\theta})(e^{2i\phi}+e^{-2i\phi})\hat{\theta}+2i(e^{2i\phi}-e^{-2i\phi})\hat{\phi}\right].
\end{eqnarray}
Again, this is the result of a ``+" polarized BWM of amplitude $h_M$ moving in the positive $z$ direction. The term proportional to ${\bf V}_\oplus$ describes a prompt deflection produced as the memory wave front passes over the observer. The term proportional to ${\bf V}_\bigstar$ describes secular evolution in the deflection angle over a time as long as $2d/c$ depending on the location of the deflected light source relative to the BWM source. 
\section{III. Temperature of the CMB}
The CMB resembles a near-perfect black body along each line of sight with a temperature of approximately 2.73 K. But the temperature varies slightly as a function of sky direction $\hat{\bf n}$. We define a ``primary" observed temperature pattern $T^0(\hat{\bf n})$. For our purposes, ``primary" simply means ``not yet perturbed by a BWM." We assume that $T^0(\hat{\bf n})$ is constant in time, though in reality it evolves over cosmological time. We decompose this primary pattern of temperature fluctuations as a linear combination of spherical harmonics:
\begin{eqnarray}
T^0(\hat{\bf n})&=&\sum_{l=0}^\infty\sum_{m=-l}^la_{lm}^0Y_l^m(\hat{\bf n}),
\end{eqnarray}
where
\begin{eqnarray}
\label{eq:primaryExpansion}
a_{lm}^0&=&\int T^0(\hat{\bf n})Y_l^{m*}(\hat{\bf n})d\Omega.
\end{eqnarray}
It will often prove useful for us in this work to express the spherical harmonics as the products of associated Legendre functions, $P_l^m$, and complex exponentials that they are
\begin{eqnarray}
Y_l^m(\theta,\phi)&=&\sqrt{\frac{(2l+1)}{4\pi}}\gamma(l,m)e^{im\phi}P_l^m(\cos{\theta}),
\end{eqnarray}
where we have defined
\begin{eqnarray}
\gamma(l,m)&=&\sqrt{\frac{(l-m)!}{(l+m)!}}.
\end{eqnarray}
We use the notations $Y_l^m(\hat{\bf n})$ and $Y_l^m(\theta,\phi)$ interchangeably and sometimes drop the argument altogether. Since $T^0(\hat{\bf n})$ is real valued, the parity properties of spherical harmonics---that $Y_l^{m*}=(-1)^mY_l^{-m}$---demand that $a^{0*}_{lm}=(-1)^ma^0_{l,-m}$. The average sky temperature can be readily expressed in terms of the monopole of this expansion, e.g. $a_{00}^0/(4\pi)^{1/2}\approx2.73$~K.

The primary temperature fluctuation pattern is usually assumed to be spatially isotropic, at least in a statistical sense. In the language of this spherical harmonic decomposition, this means that
\begin{eqnarray}
\label{eq:ensembleAverage}
\langle a^0_{l_1m_1}a^{0*}_{l_2m_2}\rangle&=&{\cal C}^0_{l_1}\delta_{l_1l_2}\delta_{m_1m_2},
\end{eqnarray}
where the angled brackets imply an ensemble average. An ensemble average is physically unrealizable since there is only one instance of the CMB in nature. But, different values of $m$ for a certain $l$ can be treated as statistically independent in an isotropic universe, so a finite average over the various $m$ values offers a useful estimator for ${\cal C}^0_l$. We call this finite average estimator
\begin{eqnarray}
\widetilde{\cal C}^0_l&=&\frac{1}{(2l+1)}\sum_{m=-l}^la^0_{lm}a^{0*}_{lm}.
\end{eqnarray}
We use the tilde to emphasize that this is a finite estimator for the ensemble average quantity ${\cal C}^0_l$. This estimator suffers from inescapable ``noise" due to the finite number of modes in a particular order $l$. This noise scales as $(2l+1)^{-1/2}$ and is commonly referred to as cosmic variance.

From the physics underlying the CMB, $\widetilde{\cal C}^0_l$ scales as $1/[l(l+1)]$ for small values of $l$. To offset this scaling, a related quantity is often studied:
\begin{eqnarray}
\widetilde{\cal D}^0_l&=&\frac{l(l+1)}{2\pi}\widetilde{\cal C}^0_l.
\end{eqnarray}
It is $\widetilde{\cal D}^0_l$ that is usually referred to as the CMB temperature power spectrum. See, e.g., Fig. 13 from a recent paper from the Atacama Cosmology Telescope Collaboration \cite{ACT20} or Fig. 57 (and others) from the Planck Collaboration \cite{Planck18}.

It is common practice to subtract out and rescale by the monopole before carrying out a spherical harmonic decomposition of the CMB temperature fluctuations. We do not do this because, as we will show, a BWM mixes power from the monopole into moments of the spatial temperature fluctuation with higher degree $l$. When one rescales the temperature fluctuations by the monopole value, the coefficients $a^0_{lm}$ are dimensionless. Since we do not do this, our expansion coefficients have dimensions of temperature. For this discussion, we have adapted the treatment of this material from the text by Maggiore \cite{Maggiore} to our purposes.
\section{IV. Manifestations of a BWM in the CMB}
  
As we discussed, a BWM induces time-variable patterns of redshifts and angular deflections that are proportional to the memory amplitude $h_M$. They prevent us from directly measuring the primary temperature fluctuation pattern of the CMB. Instead of $T^0(\hat{\bf n})$, we observe
\begin{eqnarray}
T(\hat{\bf n})&=&\frac{T^0(\hat{\bf n}+\delta{\bf n}(t,\hat{\bf n}))}{(1+z(t,\hat{\bf n}))}.
\end{eqnarray}
We expand this expression to linear order in the memory amplitude:
\begin{eqnarray}
T(\hat{\bf n})&=&T^0(\hat{\bf n})-z(t,\hat{\bf n})T^0(\hat{\bf n})+\delta{\bf n}(t,\hat{\bf n})\cdot r\nabla T^0(\hat{\bf n})+{\cal O}(h_M^2),\nonumber\\
&=&T^0(\hat{\bf n})+\delta T^\parallel(t,\hat{\bf n})+\delta T^\perp(t,\hat{\bf n})+{\cal O}(h_M^2),
\end{eqnarray}
where we have defined
\begin{eqnarray}
\delta T^\parallel(t,\hat{\bf n})&=&-z(t,\hat{\bf n})T^0(\hat{\bf n}),~~{\rm and}\\
\delta T^\perp(t,\hat{\bf n})&=&\delta{\bf n}(t,\hat{\bf n})\cdot r\nabla T^0(\hat{\bf n}).
\end{eqnarray}
We then decompose these perturbations into spherical harmonics as
\begin{eqnarray}
\delta T^\vee(t,\hat{\bf n})&=&\sum_{l=0}^\infty\sum_{m=-l}^l\delta a^\vee_{lm}(t)Y_l^m(\hat{\bf n}),
\label{eq:tempSpectrum}
\end{eqnarray}
where
\begin{eqnarray}
\delta a_{lm}^\vee(t)&=&\int \delta T^\vee(t,\hat{\bf n})Y_l^{m*}(\hat{\bf n})d\Omega.
\end{eqnarray}
We have introduced the superscript ``$\vee$" as a placeholder for either $\parallel$ or $\perp$.

Analytic calculations of the perturbations $\delta a^\vee_{lm}(t)$ are the key results of this work. These perturbations are directly proportional to the amplitude of a BWM, $h_M$, and they enter at linear order in the power spectrum of CMB temperature fluctuations:
\begin{eqnarray}
\widetilde{\cal D}_l(t)&=&\frac{1}{2\pi}\frac{l(l+1)}{(2l+1)}\sum_{m=-l}^l[a_{lm}^0+\delta a^\parallel_{lm}(t)+\delta a^\perp_{lm}(t)][a_{lm}^{0}+\delta a^\parallel_{lm}(t)+\delta a^\perp_{lm}(t)]^*+{\cal O}(h_M^2),\nonumber\\
&=&\widetilde{\cal D}_l^0+\delta\widetilde{\cal D}^\parallel_l(t)+\delta\widetilde{\cal D}^\perp_l(t)+{\cal O}(h_M^2),
\end{eqnarray}
where
\begin{eqnarray}
\delta\widetilde{\cal D}^\vee_l(t)&=&\frac{1}{2\pi}\frac{l(l+1)}{(2l+1)}\sum_{m=-l}^l a_{lm}^0\delta a^{\vee*}_{lm}(t)+a_{lm}^{0*}\delta a^{\vee}_{lm}(t).
\end{eqnarray}

\subsection{A. Temperature perturbation from memory-induced redshifts}
We now turn our focus to computing $\delta a^\parallel_{lm}(t)$. It is useful to write 
\begin{eqnarray}
z(t,\hat{\bf n})&=&\sum_{l=0}^\infty\sum_{m=-l}^l\omega_{lm}(t)Y_l^m(\theta,\phi),
\end{eqnarray}
where 
\begin{eqnarray}
\omega_{lm}(t)&=&\int z(t,\hat{\bf n})Y_l^{m*}(\hat{\bf n})d\Omega.
\end{eqnarray}
From the form of the redshift pattern in Eq. (\ref{eq:redshift}), it is clear that integration over the azimuthal coordinate $\phi$ is trivial, that $\omega_{lm}$ is nonzero only when $m=\pm2$, and that $\omega_{l,+2}=\omega_{l,-2}$. Focusing on the case where $m=+2$ and doing the integration over $\phi$, we are left with
\begin{eqnarray}
\omega_{l,+2}(t)=\frac{\pi h_M }{2}\sqrt{\frac{(2l+1)}{4\pi}}\gamma(l,+2)\int_0^\pi\sin{\theta}(1-\cos{\theta})\Theta(\theta_t-\theta)P_l^{+2}(\cos{\theta})d\theta,
\end{eqnarray}
where, as a reminder, $\cos{\theta_t}=\beta-1$, $\beta=ct/d$, and we limit our focus to $0\leq\beta\leq2$. The effect of the Heaviside step function is to truncate the integration interval. After changing the integration variable to $x=\cos{\theta}$, this becomes
\begin{eqnarray}
\omega_{l,+2}(t)&=&\frac{\pi h_M}{2}\sqrt{\frac{(2l+1)}{4\pi}}\gamma(l,+2)\bigg[{\cal I}_{G,l}\left(\beta-1,1\right)-{\cal I}_{H,l}\left(\beta-1,1\right)\bigg]~~{\rm for}~l\geq2,
\end{eqnarray}
where ${\cal I}_{G,l}$ and ${\cal I}_{H,l}$ are given explicitly in Appendix B---they are straightforward combinations of associated Legendre functions. 

Inserting this decomposition of the BWM-induced redshift pattern and the decomposition of $T^0(\hat{\bf n})$ from Eq. (\ref{eq:primaryExpansion}) into our expression for $\delta a_{lm}^\parallel(t)$ yields
\begin{eqnarray}
\delta a_{l_1m_1}^\parallel(t)&=&-\sum_{l_2=0}^\infty\sum_{m_2=-l_2}^{l_2}a_{l_2m_2}^0\sum_{l_3=2}^\infty\omega_{l_3,+2}(t)\int[Y_{l_3}^{-2}(\hat{\bf n})+Y_{l_3}^{+2}(\hat{\bf n})]Y_{l_2}^{m_2}(\hat{\bf n})Y_{l_1}^{m_1*}(\hat{\bf n})d\Omega.
\end{eqnarray}
Integrals over the sphere of products of three spherical harmonics can be written in terms of Clebsch-Gordan coefficients:
\begin{eqnarray}
\label{eq:tripleSpherical}
\int Y_{l_3}^{m_3}(\hat{\bf n})Y_{l_2}^{m_2}(\hat{\bf n})Y_{l_1}^{m_1*}(\hat{\bf n})d\Omega&=&\sqrt{\frac{(2l_2+1)(2l_3+1)}{4\pi(2l_1+1)}}~C_{l_2,0;l_3;0}^{l_1,0}C_{l_2,m_2;l_3,m_3}^{l_1,m_1}.
\end{eqnarray}
The Clebsch-Gordan coefficients are symmetric under interchange of the lower two pairs of indices, i.e., $C_{l_2,m_2;l_3,m_3}^{l_1,m_1}=C_{l_3,m_3;l_2,m_2}^{l_1,m_1}$. ``Bra-ket" notation is often used for Clebsch-Gordan coefficients, i.e.,
\begin{eqnarray}
C_{l_2,m_2;l_3,m_3}^{l_1,m_1}&=&\langle l_2,m_2,l_3,m_3|l_1,m_1\rangle.
\end{eqnarray}
We avoid bra-ket notation for compactness and to facilitate related notation introduced in Appendix C. For a Clebsch-Gordan coefficient to not vanish, the ``$l$" values must satisfy a triangle inequality: $|l_2-l_3|\leq l_1\leq l_2+l_3$. Additionally, the Clebsch-Gordan coefficients vanish unless $m_1=m_2+m_3$, enforcing the selection rule that arises from carrying out the integral over $\phi$. In the context of our expression for $\delta a_{lm}^\parallel(t)$, this fact allows us to eliminate the summation over $m_2$:
\begin{eqnarray}
\label{eq:redshiftPert}
\delta a_{l_1m_1}^\parallel(t)&=&-\sum_{l_2=0}^\infty\sum_{l_3=2}^\infty\omega_{l_3,+2}(t)\sqrt{\frac{(2l_2+1)(2l_3+1)}{4\pi(2l_1+1)}}C_{l_2,0;l_3,0}^{l_1,0}\left[a^0_{l_2,m_1+2}C_{l_2,m_1+2;l_3,-2}^{l_1,m_1}+a^0_{l_2,m_1-2}C_{l_2,m_1-2;l_3,+2}^{l_1,m_1}\right].
\end{eqnarray}
To compute $\delta a^\perp_{lm}(t)$, we follow very much the same steps followed here, but the calculation is sufficiently more complicated that we reserve the details for the appendixes.

\section{V. Discussion}

The framework we have developed can be used to compute $\delta a_{lm}^\vee(t)$, $\delta T^\vee(t,\hat{\bf n})$, and $\delta\widetilde{\cal D}_l^\vee(t)$, for any values of $a_{lm}^0$ out to arbitrarily large values of $l$. In practice, the infinite summations in Eqs. (\ref{eq:tempSpectrum}), (\ref{eq:redshiftPert}), and (\ref{eq:deflectionPert}) need to be truncated at some value of $l$, $l_{\rm trunc}$. Important structure in the CMB temperature power spectrum from baryon acoustic oscillations extends up to values of $l\approx2000$, so any reasonable choice for $l_{\rm trunc}$ for real-world applications should exceed 2000. In that case, Eqs. (\ref{eq:redshiftPert}) and (\ref{eq:deflectionPert}) involve the calculation of several million terms. Many of those terms vanish. Judicious exploitation of the triangle inequalities that Clebsch-Gordan coefficients must satisfy in order to not vanish can greatly reduce the number of terms that actually need to be computed. But even without exploiting these triangle inequalities, this is not an especially onerous numerical problem. We have analytically carried out all of the difficult integration and reduced the problem to computation of Clebsch-Gordan coefficients and evaluation of associated Legendre functions, both things with efficient numerical implementations. Nonetheless, we reserve a detailed application of these tools to the full CMB power spectrum and the phenomenological investigation such a project merits for later work. 

\begin{figure}[]
    \includegraphics[scale=.7]{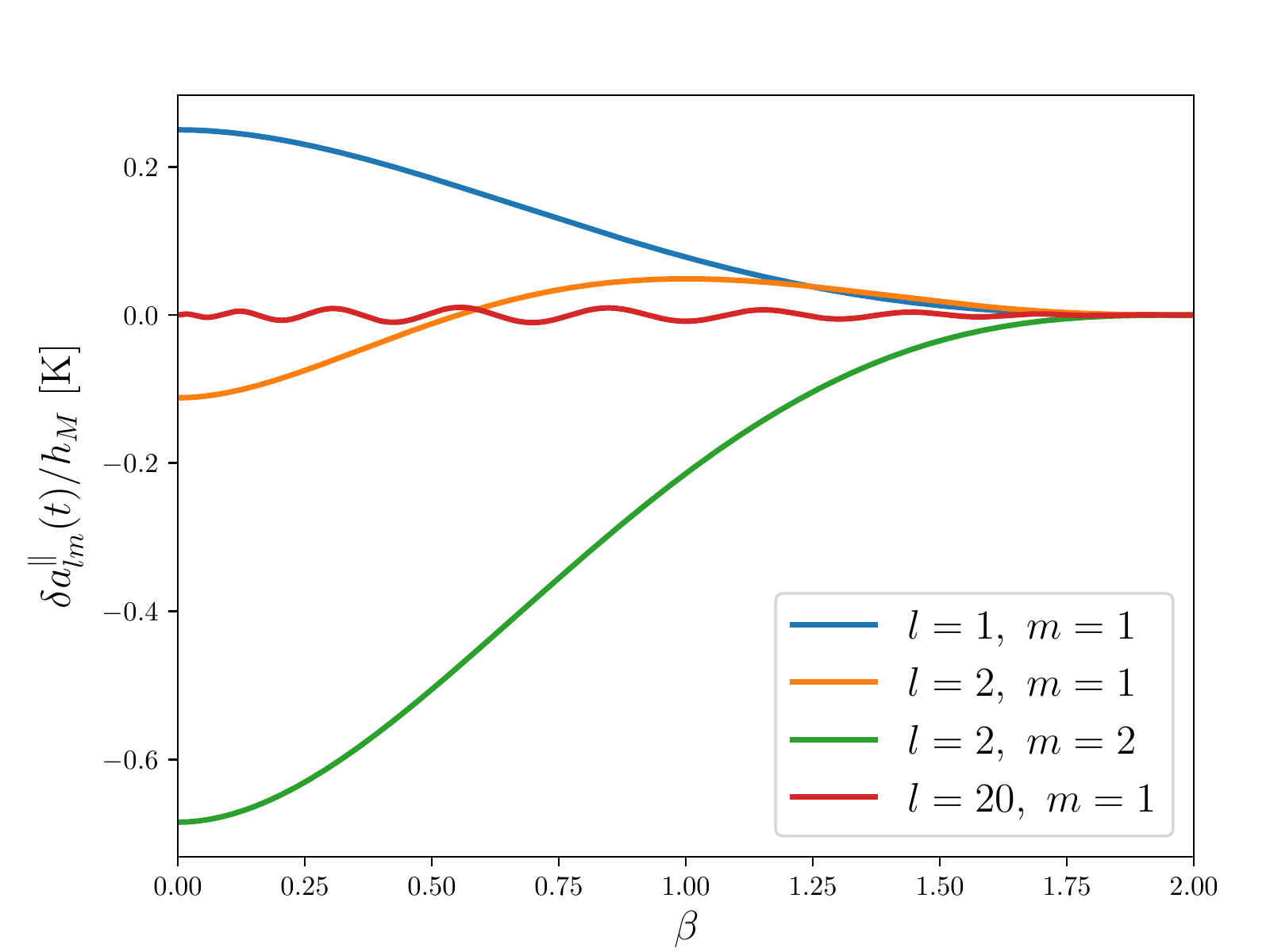}
    \caption{Multipolar perturbations from BWM-induced redshifts versus time for a small sample of $l$ and $m$ values. We started with a toy version of the CMB primary temperature pattern where $a_{0,0}^0=3$K, $a_{1,-1}^0=-a_{1,1}^0=1$K, and all other values of $a_{lm}^0=0$. These curves were all computed using Eq. (\ref{eq:redshiftPert}). The dimensionless time variable $\beta=ct/d$ ranges from 0, when the BWM wave front first encounters the observer, to 2, twice the light travel time to the source of light being observed---since we are concerned with the CMB, $d$ is the distance to the surface of last scattering. All of the redshift perturbations go to zero as $\beta$ goes to 2. Even with this toy CMB primary where all power is confined to $l\leq1$, BWM-induced redshifts produce perturbations of some amplitude up to arbitrarily-large values of $l$---we include the $l=20$, $m=1$ curve to indicate this. The curves with $m=1$ are nonzero because of coupling to the dipole of the CMB. The curve with $m=2$ is nonzero because of coupling to the monopole.}
    \label{fig:deltaAlmParallel}
\end{figure}

\begin{figure}[]
    \includegraphics[scale=.7]{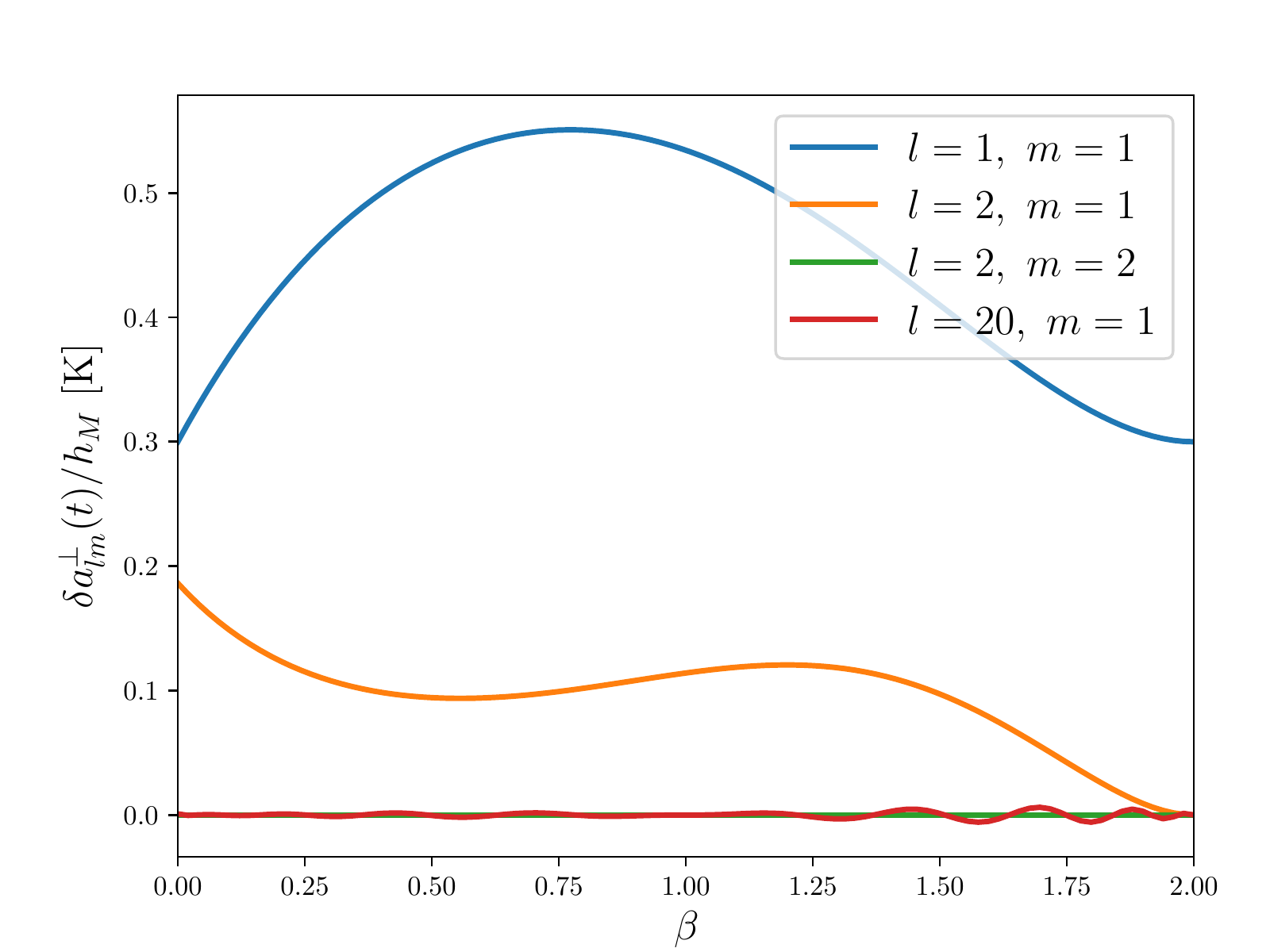}
    \caption{Multipolar perturbations from BWM-induced deflections with a toy CMB primary identical to what was used in Fig. \ref{fig:deltaAlmParallel}. These curves were all computed using Eq. (\ref{eq:deflectionPert}). Unlike the redshift-induced perturbations from Fig. \ref{fig:deltaAlmParallel}, these deflection-induced perturbations do not all go to zero as $\beta$ goes to 2, indicating the permanence of BWM deflection distortions discussed in \cite{m20}. In this case, the $l=2$, $m=2$ curve is identically zero because unlike the redshift perturbations, the deflection perturbations do not couple to the CMB's monopole.}
    \label{fig:deltaAlmPerp}
\end{figure}

Here, we consider greatly simplified toy models of the CMB primary that still allow us to straightforwardly demonstrate many of the implications of the analysis we have done. First, we consider a CMB where $a_{0,0}^0=3{\rm K}$ and $a_{1,-1}^0=-a_{1,1}^0=1{\rm K}$---otherwise, $a_{lm}^0=0$. This is a sky where the temperature is real and positive everywhere. The primary has a dipolar asymmetry but no additional structure. With this simple primary, we plot $\delta a_{lm}^\parallel(t)$ in Fig. 1 and $\delta a_{lm}^\perp(t)$ in Fig. 2 for a selection of $l$ and $m$ values. The first nonvanishing perturbations are $\delta a_{1,1}^\vee(t)$ shown in blue in Figs. 1 and 2 [note that $\delta a_{l,-1}^\vee(t)=-\delta a_{l,1}^\vee(t)$]. Perturbations with $m=\pm1$ (see the blue, orange, and red curves in Figs. 1 and 2) are nonzero because of terms in the summations of Eqs.~(\ref{eq:redshiftPert}) and (\ref{eq:deflectionPert}) proportional to $a_{1,\mp 1}^0$. This means power in the dipole is being coupled to modes with $l\geq 1$. Perturbations with arbitrarily large values of $l$ can be nonzero (see the red curves in Figs. 1 and 2 associated with $l=20$) though the amplitude of the perturbation falls with increasing $l$. This pumping of power into high values of $l$ is tied, at least in part, to the sharp transition edges from the step functions appearing in Eqs. (\ref{eq:redshift}) and (\ref{eq:deflection}). For CMB surveys that focus on a particular window of the sky rather than the full sphere (like the one done by the Atacama Cosmology Telescope \cite{ACT20}) the transition edge for a particular BWM may not appear in the window being observed. Additional steps must be applied to our framework in order to directly apply it to such windowed surveys. The green curve in Fig. 1 depicting $\delta a_{2,2}^\parallel(t)$ is nonzero because of terms in Eq. (\ref{eq:redshiftPert}) proportional to $a_{0,0}^0$, i.e. this perturbation is sourced by the monopole of the primary. Such terms never appear in Eq. (\ref{eq:deflectionPert}) so $\delta a_{2,2}^\perp(t)=0$. An additional stark difference between Figs. 1 and 2 is that unlike in Fig. 2, all of the curves in Fig. 1 go to zero as $\beta$ goes to 2. This speaks to the fact that BWM-induced redshift perturbations do eventually vanish completely, but deflection perturbations are actually persistent \cite{m20}. 
With this simple toy primary CMB, it is straightforward to compute $\delta\widetilde{\cal D}_l^\vee(t)$. Because we have set $a_{lm}^0=0$ for all $l>1$, $\delta\widetilde{\cal D}_l^\vee$ can only be nonzero (or, more accurately, linear in $h_M$---tiny quadratic contributions that we ignore may be present) for $l=0$ or $1$. Since $\delta a_{0,0}^\vee(t)=0$, $\delta\widetilde{\cal D}_0^\vee(t)=0$. We can then show that $\delta\widetilde{\cal D}_1^\vee(t)=(4/3\pi)\delta a_{1,1}^\vee(t)a_{1,1}^0$, i.e. the redshift and deflection perturbations to the dipolar power are proportional to the blue curves in Figs. \ref{fig:deltaAlmParallel} and \ref{fig:deltaAlmPerp} respectively.

\begin{figure}[]
    \subfloat[]{\includegraphics[scale=.59]{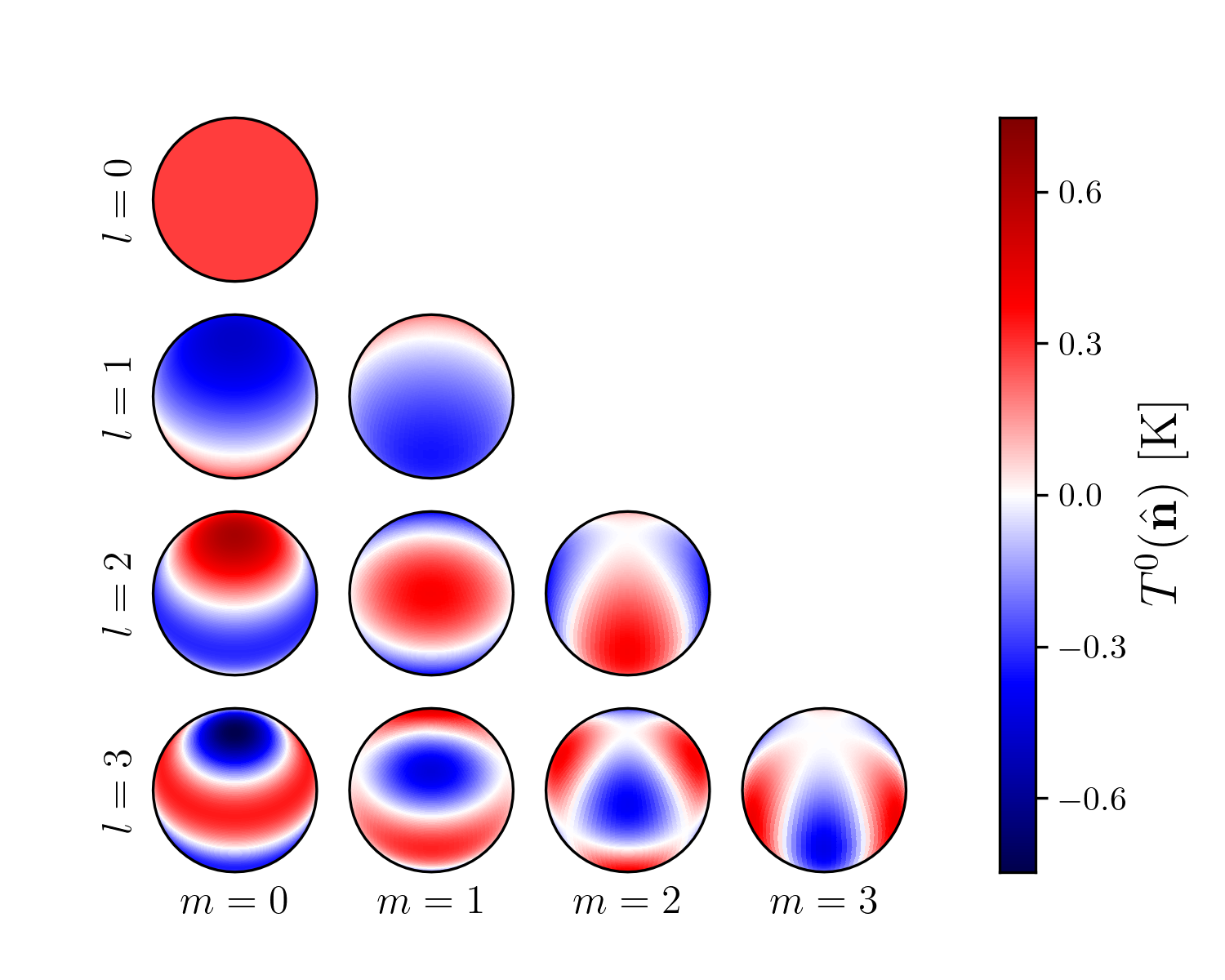}}\\
    \subfloat[]{\includegraphics[scale=.59]{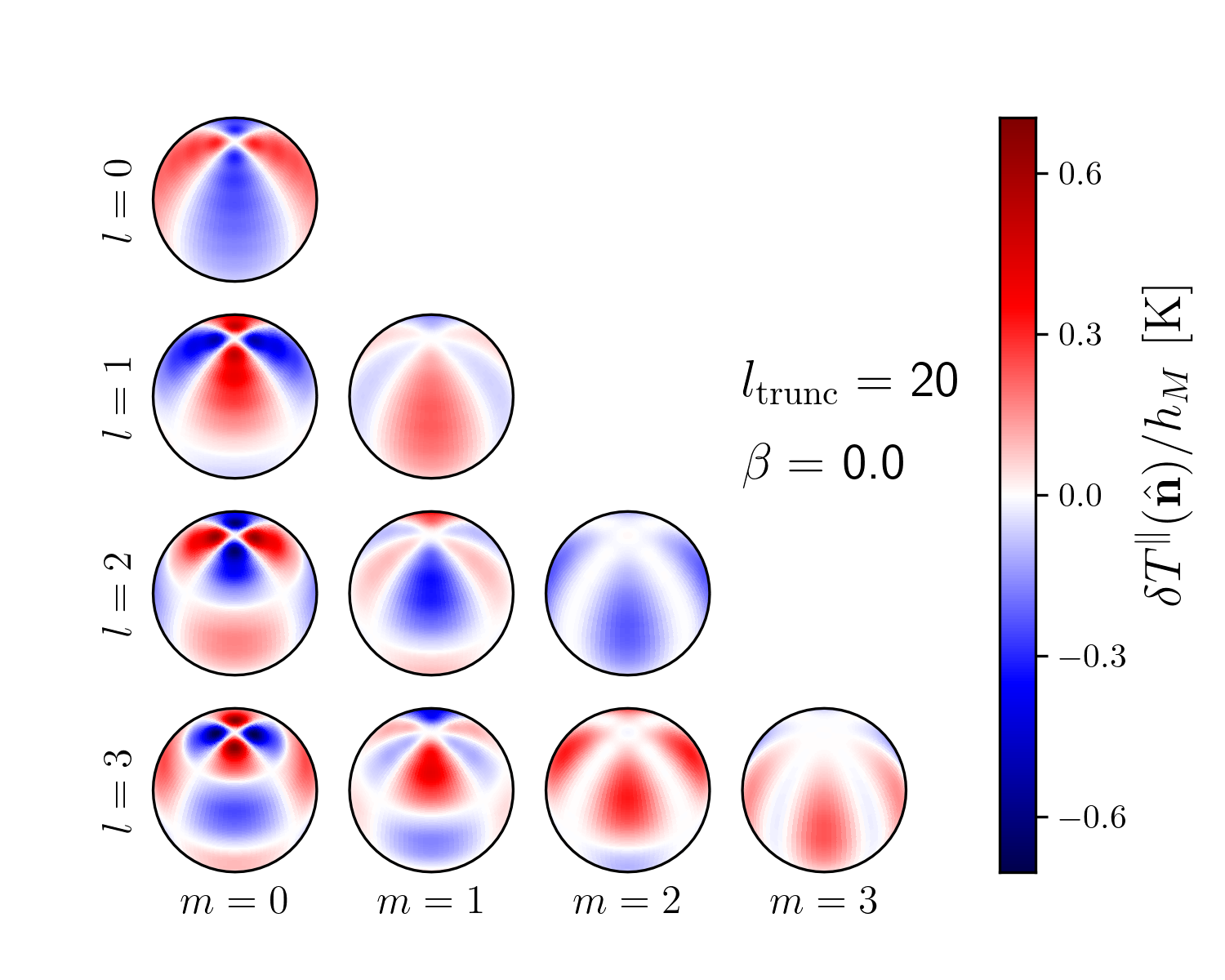}}
    \subfloat[]{\includegraphics[scale=.59]{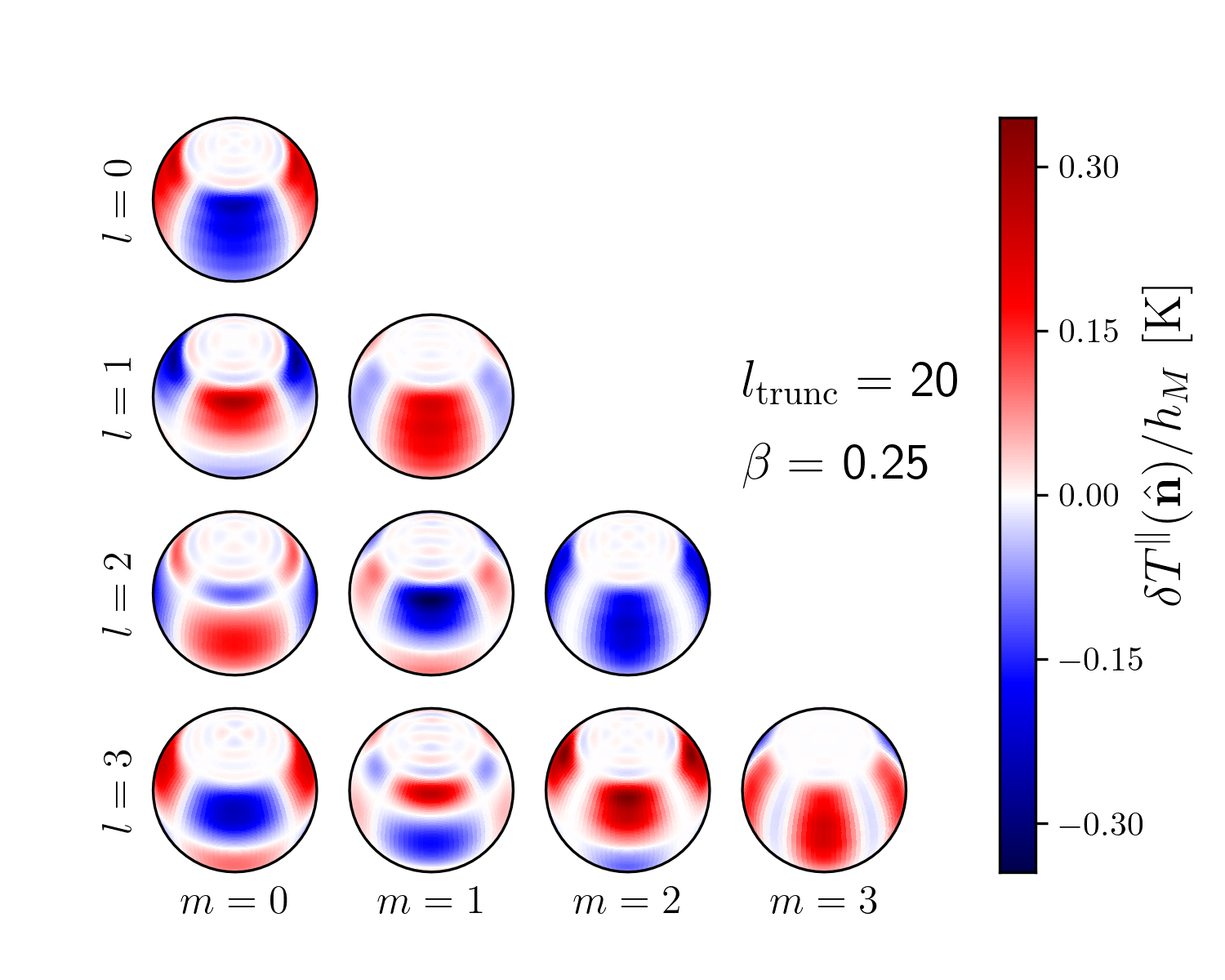}}\\
    \subfloat[]{\includegraphics[scale=.59]{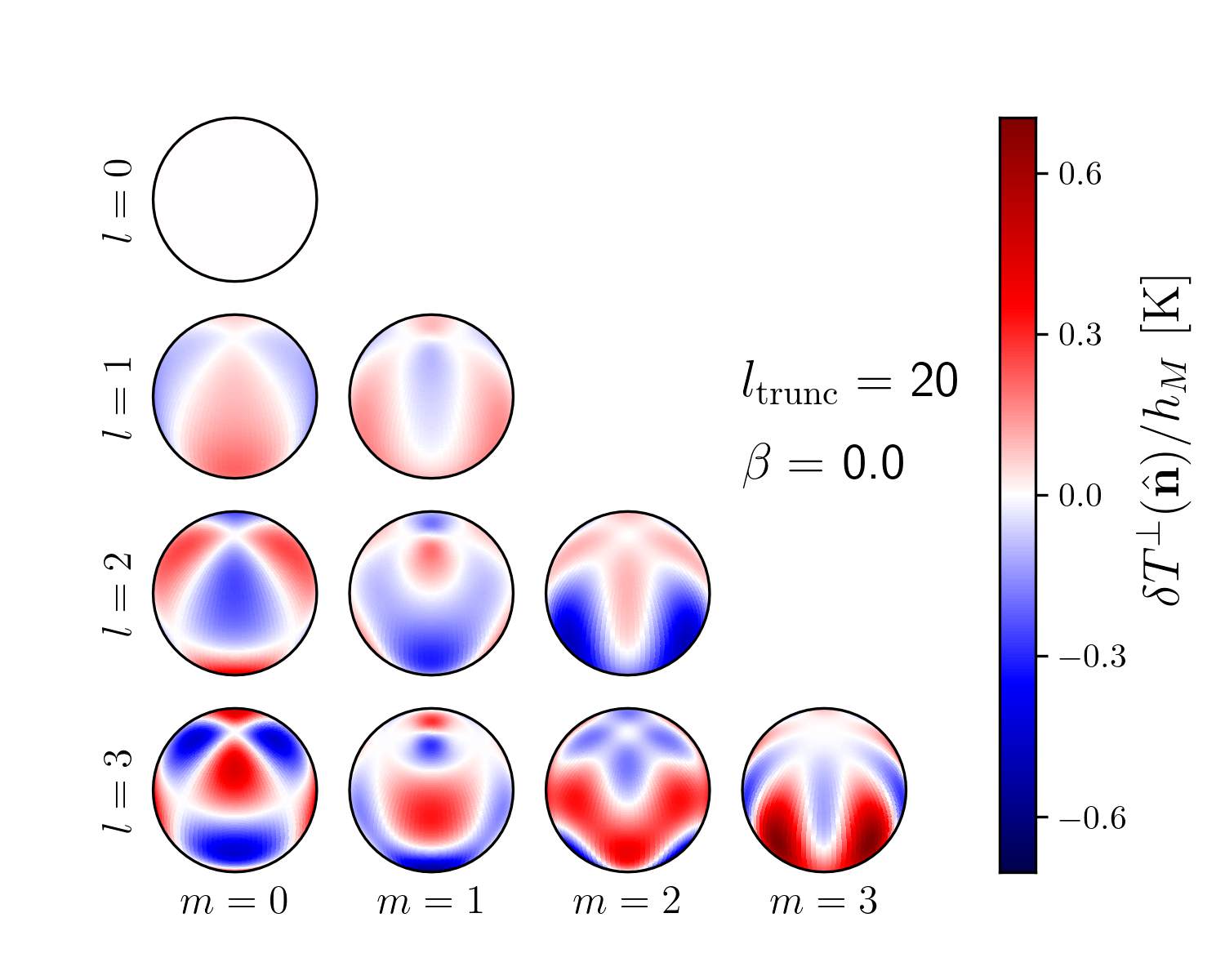}}
    \subfloat[]{\includegraphics[scale=.59]{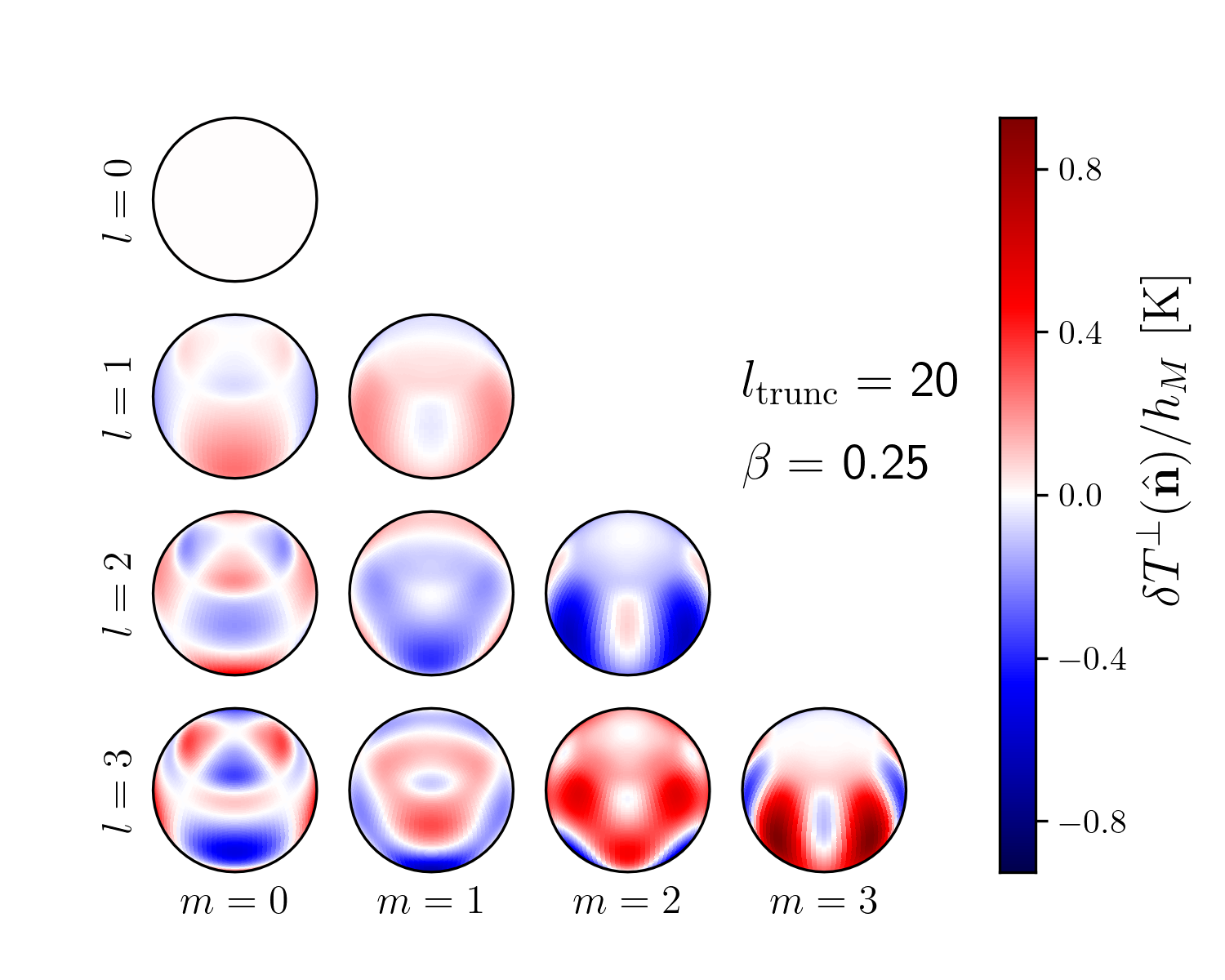}}
     \caption{(a) Different temperature patterns equal to the real-valued spherical harmonics (i.e., $Y_l^m$ for $m=0$ and $[Y_l^m+(-1)^mY_l^{-m}]/2$ for $m>0$) that may contribute to the primary CMB temperature pattern. (b), (c) Temperature perturbations from redshifts (divided by the memory amplitude $h_M$) to the spherical harmonic modes from (a) caused by a BWM propagating into the displayed southern pole at times $\beta=0.0$ and $\beta=0.25$, respectively. We computed these using Eqs. (\ref{eq:tempSpectrum}) and (\ref{eq:redshiftPert}) and truncated the infinite summations therein at $l_{\rm trunc}=20$. (d), (e) Perturbations from BWM-induced deflections to the spherical harmonic modes of the primary at times $\beta=0.0$ and $\beta=0.25$, respectively. These were computed using Eqs. (\ref{eq:tempSpectrum}) and (\ref{eq:deflectionPert}) with the infinite summations again truncated at $l_{\rm trunc}=20$.}
    \label{fig:deflectionPatterns}
\end{figure}

Figures \ref{fig:deltaAlmParallel} and \ref{fig:deltaAlmPerp} demonstrate the time evolution of various BWM-induced perturbations $\delta a^\vee_{lm}(t)$ for a specific, simple primary CMB temperature pattern. As a second demonstration of this formalism, in Fig. \ref{fig:deflectionPatterns} we show fixed time snapshots of the spatial temperature perturbation patterns $\delta T^\vee(t,\hat{\bf n})$ produced by a BWM acting on different spherical harmonic contributions to the CMB temperature primary. Figure 3(a) shows real-valued versions of the spherical harmonics representing possible contributions to the CMB primary: $Y_l^m$ when $m=0$ and $[Y_l^m+(-1)^mY_l^{-m}]/2$ when $m>0$. Different rows are associated with different values of $l$ and different columns are associated with different values of $m$ as labeled. The spherical projections have been oriented so as to improve the reader's view of the southern pole into which a BWM is propagating in our formalism. Figures 3(b) and 3(c) depict the BWM-induced redshift perturbation to the primary CMB temperature pattern, $\delta T^\parallel(\hat{\bf n})$, at values of time $\beta=0.0$ and $\beta=0.25$, respectively. We divided out the memory amplitude, $h_M$. We used Eqs. (\ref{eq:tempSpectrum}) and (\ref{eq:redshiftPert}) to compute these patterns. Looking to Figs. \ref{fig:deltaAlmParallel} and \ref{fig:deltaAlmPerp} where the scale of $\delta a_{lm}^\vee$ is much smaller at $l=20$ than at lower values of $l$, we have truncated the infinite summations in Eqs. (\ref{eq:tempSpectrum}) and (\ref{eq:redshiftPert}) at $l_{\rm trunc}=20$ for computational expediency. Of particular note, a monopolar contribution to the CMB primary leads to nonvanishing values of $\delta T^\parallel(\hat{\bf n})$ as indicated by the uppermost sphere in Figs. 3(b) and 3(c). As time progresses from $\beta=0.0$ to $\beta=0.25$, a polar cap develops around the direction from which the BWM came where the redshift perturbations turn off. Lingering low-amplitude structure visible in this polar cap region results from our truncation of the infinite summations. Figures~3(d) and 3(e) are the same as 3(b) and 3(c) but for the the BWM-induced deflection perturbation to the primary CMB temperature pattern, $\delta T^\perp(\hat{\bf n})$. Notably, when the primary CMB temperature pattern is just a monopole, the deflection perturbation vanishes. As time progresses from $\beta=0.0$ to $\beta=0.25$, the deflection perturbation does not vanish in an expanding polar cap---the deflection perturbations persist.

Possible next steps for this line of research are myriad. We have mentioned that this framework should be applied to realistic models of the CMB primary and the influence to the temperature power spectrum should be scrutinized. However, before such work is belabored, or perhaps in parallel with such work, the way in which spherical wave fronts influence the results presented here should be investigated; we have started to look at this with colleagues. We should compare the effects of BWMs on the CMB with other phenomena that undermine our ability to probe the true primordial CMB primary---things like dust extinction and lensing from foreground structure. One may wish to extend this type of analysis to non-Einsteinian types of memory with nontransverse polarizations that may arise in modified theories of gravity \cite{dn16,k20}; this may require revisiting some of the work of Book and Flanagan \cite{bf11} that our results are built upon. Ultimately, since the effects of BWMs are so long-lasting, we will care about the superposition of redshift and deflection effects from a cosmological population of BWM sources occurring all over the sky and at a range of times in the past. This will prove to be a rich line of enquiry that is necessary to assess the detectability of BWM signatures in the CMB. The entire GW history of the Universe contributes to the signal we are pursuing.

\section{Appendix A: Vector Spherical Harmonics}
For the calculations in these appendixes we make extensive use of so-called ``vector spherical harmonics." There are a variety of conventions in use for them, so here we lay out in detail the conventions we use. First, we define the angular momentum operator ${\bf L}=-i{\bf r}\times\nabla$, a vector differential operator that acts on functions of the angular spherical coordinates. For application to spherical harmonics $Y_l^m(\theta,\phi)$, a useful form of the angular momentum operator is
\begin{eqnarray}
{\bf L}&=&\bigg[\frac{1}{2}\cos{\theta}(e^{-i\phi}L_++e^{i\phi}L_-)-\sin{\theta}L_z\bigg]\hat{\theta}-\frac{i}{2}(e^{-i\phi}L_+-e^{i\phi}L_-)\hat{\phi},
\end{eqnarray}
where the operators $L_\pm$ and $L_z$ act on spherical harmonics as
\begin{eqnarray}
L_\pm Y_l^m(\theta,\phi)&=&\Lambda_\pm(l,m)Y_l^{m\pm1}(\theta,\phi),~{\rm and}\\
L_z Y_l^m(\theta,\phi)&=&m Y_l^m(\theta,\phi).
\end{eqnarray}
We introduced the coefficients
\begin{eqnarray}
\Lambda_\pm(l,m)&=&\sqrt{(l\mp m)(l\pm m+1)}.
\end{eqnarray}
Next, we define another operator ${\bf K}=\hat{\bf r}\times{\bf L}=ir\nabla$. With these two operators, we define the vector spherical harmonics:
\begin{eqnarray}
\label{eq:Philm}
{\bf \Phi}_l^m(\theta,\phi)&=&{\bf L}Y_l^m(\theta,\phi),\nonumber\\
&=&\bigg[\frac{1}{2}\cos{\theta}\bigg(\Lambda_+(l,m)e^{-i\phi}Y_l^{m+1}(\theta,\phi)+\Lambda_-(l,m)e^{i\phi}Y_l^{m-1}(\theta,\phi)\bigg)-m\sin{\theta}Y_l^m(\theta,\phi)\bigg]\hat{\theta}-\nonumber\\
&&\frac{i}{2}\bigg(\Lambda_+(l,m)e^{-i\phi}Y_l^{m+1}(\theta,\phi)-\Lambda_-(l,m)e^{i\phi}Y_l^{m-1}(\theta,\phi)\bigg)\hat{\phi},~{\rm and}\\
\label{eq:Psilm}
{\bf \Psi}_l^m(\theta,\phi)&=&{\bf K}Y_l^m(\theta,\phi),\nonumber\\
&=&\frac{i}{2}\bigg(\Lambda_+(l,m)e^{-i\phi}Y_l^{m+1}(\theta,\phi)-\Lambda_-(l,m)e^{i\phi}Y_l^{m-1}(\theta,\phi)\bigg)\hat{\theta}+\nonumber\\
&&\bigg[\frac{1}{2}\cos{\theta}\bigg(\Lambda_+(l,m)e^{-i\phi}Y_l^{m+1}(\theta,\phi)+\Lambda_-(l,m)e^{i\phi}Y_l^{m-1}(\theta,\phi)\bigg)-m\sin{\theta}Y_l^m(\theta,\phi)\bigg]\hat{\phi}
\end{eqnarray}

The two types of vector spherical harmonics are spatially orthogonal when they are of the same degree $l$ and order $m$, i.e., ${\bf \Phi}_{l_1}^{m_1}\cdot{\bf \Psi}_{l_2}^{m_2}=0$ if $l_1=l_2$ and $m_1=m_2$. Under integration over the sphere, they have additional orthogonality properties:
\begin{eqnarray}
\int{\bf \Phi}_{l_1}^{m_1}\cdot{\bf \Psi}_{l_2}^{m_2*}d\Omega&=&0,\\
\int{\bf \Phi}_{l_1}^{m_1}\cdot{\bf \Phi}_{l_2}^{m_2*}d\Omega&=&l_1(l_1+1)\delta_{l_1l_2}\delta_{m_1m_2},\\
\int{\bf \Psi}_{l_1}^{m_1}\cdot{\bf \Psi}_{l_2}^{m_2*}d\Omega&=&l_1(l_1+1)\delta_{l_1l_2}\delta_{m_1m_2}.
\end{eqnarray}
It is these orthogonality properties that make the vector spherical harmonics a convenient basis for multipolar decompositions of vector fields on the sphere.
\section{Appendix B: Useful Identities and Integrals Involving Associated Legendre Functions}

Much of the mathematical work in this paper boils down to carrying out integrals involving associated Legendre functions. Here, we list a few identities that allowed us to carry out the integrals we encountered and explicitly write out the results of the integrals used in this paper. First, two useful integral identities:
\begin{eqnarray}
\int_a^b(1-x^2)^{-m/2}P_l^m(x)dx=-(1-x^2)^{-(m-1)/2}P_l^{m-1}(x)\bigg|_a^b,
\end{eqnarray}
and
\begin{eqnarray}
\int_a^b(1-x^2)^{m/2}P_l^m(x)dx=\frac{(1-x^2)^{(m+1)/2}}{(l-m)(l+m+1)}P_l^{m+1}\bigg|_a^b.
\end{eqnarray}
These indefinite integral identities can be found in the Digital Library for Mathematical Functions' section on Legendre functions \cite{DLMF}, specifically Sec. 14.17. All of the integrals we will face can be made to resemble one of the two integrals listed above through use of the following common recurrence relations:
\begin{eqnarray}
\label{eq:recurrence1}
xP_l^m(x)&=&\frac{1}{(2l+1)}[(l-m+1)P_{l+1}^m(x)+(l+m)P_{l-1}^m(x)],
\end{eqnarray}
and
\begin{eqnarray}
\label{eq:recurrence2}
(1-x^2)^{1/2}P_l^m(x)&=&\frac{1}{(2l+1)}[(l-m+1)(l-m+2)P_{l+1}^{m-1}(x)-(l+m-1)(l+m)P_{l-1}^{m-1}].
\end{eqnarray}

The following integrals are utilized in this work. These expressions are correct at least for all values of $l\geq1$ which is what we require for our purposes.
\begin{eqnarray}
{\cal I}_{A,l}(a,b)&=&\int_a^b(1-x^2)^{-1/2}P_l^{+1}(x)dx,\nonumber\\
&=&-P_l(x)\bigg|_a^b.
\end{eqnarray}
\begin{eqnarray}
{\cal I}_{B,l}(a,b)&=&\int_a^b(1-x^2)^{-1/2}x~P_l^{+1}(x)dx,\nonumber\\
&=&\frac{1}{(2l+1)}[l{\cal I}_{A,l+1}(a,b)+(l+1){\cal I}_{A,l-1}(a,b)].
\end{eqnarray}
\begin{eqnarray}
{\cal I}_{C,l}(a,b)&=&\int_a^b(1-x^2)^{-1/2}x^2P_l^{+1}(x)dx,\nonumber\\
&=&\frac{1}{(2l+1)}[l{\cal I}_{B,l+1}(a,b)+(l+1){\cal I}_{B,l-1}(a,b)].
\end{eqnarray}
\begin{eqnarray}
{\cal I}_{D,l}(a,b)&=&\int_a^b(1-x^2)^{1/2}P_l^{+1}(x)dx,\nonumber\\
&=&\left\{\begin{array}{c}
\frac{1}{3}x(x^2-3)\bigg|_a^b~~~~~~~~~~~~~~~~;~~~l=1\\\\
\frac{(1-x^2)}{(l-1)(l+2)}P_l^{+2}(x)\bigg|_a^b~~~~~~;~~l>1
\end{array}\right.~~~~.
\end{eqnarray}
\begin{eqnarray}
{\cal I}_{E,l}(a,b)&=&\int_a^b(1-x^2)^{1/2}x~P_l^{+1}(x)dx,\nonumber\\
&=&\frac{1}{(2l+1)}[l{\cal I}_{D,l+1}(a,b)+(l+1){\cal I}_{D,l-1}(a,b)].
\end{eqnarray}
\begin{eqnarray}
{\cal I}_{F,l}(a,b)&=&\int_a^b(1-x^2)^{1/2}x^2~P_l^{+1}(x)dx,\nonumber\\
&=&\frac{1}{(2l+1)}[l{\cal I}_{E,l+1}(a,b)+(l+1){\cal I}_{E,l-1}(a,b)].
\end{eqnarray}
\begin{eqnarray}
{\cal I}_{G,l}(a,b)&=&\int_a^bP_l^{+2}(x)dx,\nonumber\\
&=&-\frac{1}{(2l+1)}[(l-1)lP_{l+1}(x)-(l+1)(l+2)P_{l-1}(x)]\bigg|_a^b.
\end{eqnarray}
\begin{eqnarray}
{\cal I}_{H,l}(a,b)&=&\int_a^bx~P_l^{+2}(x)dx,\nonumber\\
&=&\frac{1}{(2l+1)}[(l-1){\cal I}_{G,l+1}(a,b)+(l+2){\cal I}_{G,l-1}(a,b)].
\end{eqnarray}
\begin{eqnarray}
{\cal I}_{I,l}(a,b)&=&\int_a^bx^2P_l^{+2}(x)dx,\nonumber\\
&=&\frac{1}{(2l+1)}[(l-1){\cal I}_{H,l+1}(a,b)+(l+2){\cal I}_{H,l-1}(a,b)].
\end{eqnarray}
\begin{eqnarray}
{\cal I}_{J,l}(a,b)&=&\int_a^bx^3P_l^{+2}(x)dx,\nonumber\\
&=&\frac{1}{(2l+1)}[(l-1){\cal I}_{I,l+1}(a,b)+(l+2){\cal I}_{I,l-1}(a,b)].
\end{eqnarray}
\begin{eqnarray}
{\cal I}_{K,l}(a,b)&=&\int_a^b(1-x^2)^{1/2}P_l^{+3}(x)dx,\nonumber\\
&=&\bigg[\frac{(l-2)(l-1)(l+2)(l+3)}{(2l+1)(2l+3)}+\frac{(l-2)(l-1)(l+2)(l+3)}{(2l-1)(2l+1)}\bigg]P_l(x)-\nonumber\\
&&\frac{(l-2)(l-1)l(l+1)}{(2l+1)(2l+3)}P_{l+2}(x)-\frac{l(l+1)(l+2)(l+3)}{(2l-1)(2l+1)}P_{l-2}(x)\bigg|_a^b.
\end{eqnarray}
\begin{eqnarray}
{\cal I}_{L,l}&=&\int_a^b(1-x^2)^{1/2}x~P_l^{+3}(x)dx,\nonumber\\
&=&\frac{1}{(2l+1)}[(l-2){\cal I}_{K,l+1}(a,b)+(l+3){\cal I}_{K,l-1}(a,b)].
\end{eqnarray}
\begin{eqnarray}
{\cal I}_{M,l}&=&\int_a^b(1-x^2)^{1/2}x^2P_l^{+3}(x)dx,\nonumber\\
&=&\frac{1}{(2l+1)}[(l-2){\cal I}_{L,l+1}(a,b)+(l+3){\cal I}_{L,l-1}(a,b)].
\end{eqnarray}
\begin{eqnarray}
{\cal I}_{N,l}&=&\int_a^b\frac{(1-x^2)^{1/2}}{(1+x)}P_l^{+3}(x)dx\nonumber\\
&=&-\frac{(1-x^2)^{-1/2}}{(2l+1)}\bigg\{(l-2)(l-1)P_{l+1}^{+1}(x)-(l+2)(l+3)P_{l-1}^{+1}(x)-\nonumber\\
&&~~~~~~~~~~~~~~~~~~~~~~~~~~\frac{(l-2)(l-1)}{(2l+3)}[lP_{l+2}^{+1}(x)+(l+3)P_l^{+1}(x)]+\nonumber\\
&&~~~~~~~~~~~~~~~~~~~~~~~~~~\frac{(l+2)(l+3)}{(2l-1)}[(l-2)P_l^{+1}(x)+(l+1)P_{l-2}^{+1}(x)]\bigg\}\bigg|_a^b.
\end{eqnarray}
\begin{eqnarray}
{\cal I}_{O,l}&=&\int_a^b\frac{(1-x^2)^{1/2}}{(1+x)}xP_l^{+3}(x)dx\nonumber\\
&=&\frac{1}{(2l+1)}[(l-2){\cal I}_{N,l+1}(a,b)+(l+3){\cal I}_{N,l-1}(a,b)].
\end{eqnarray}
The bounds of integration, $a$ and $b$, are assumed to be between $-1$ and $1$, inclusive. Both ${\cal I}_{N,l}$ and ${\cal I}_{O,l}$ are proportional to $(1-x^2)^{-1/2}$, so evaluating them at exactly $\pm1$ will lead to division by zero. However, evaluating them at $\pm(1-\epsilon)$ and taking the limit as $\epsilon$ goes to zero leads to well-behaved, convergent results.
\section{Appendix C: Vector Extensions of the Clebsch-Gordan Coefficients}
To complete our description of the way that a BWM-induced deflection pattern influences the CMB, we will need to know
\begin{eqnarray}
\Delta_{l_2,m_2;l_3,m_3}^{l_1,m_1}&=&\int[{\bf\Phi}_{l_3}^{m_3}({\hat{\bf n}})\cdot{\bf\Psi}_{l_2}^{m_2}(\hat{\bf n})]Y_{l_1}^{m_1*}(\hat{\bf n})d\Omega,~{\rm and}\\
\Gamma_{l_2,m_2;l_3,m_3}^{l_1,m_1}&=&\int[{\bf\Psi}_{l_3}^{m_3}({\hat{\bf n}})\cdot{\bf\Psi}_{l_2}^{m_2}(\hat{\bf n})]Y_{l_1}^{m_1*}(\hat{\bf n})d\Omega.
\end{eqnarray}
These integrals are similar in form to Eq. (\ref{eq:tripleSpherical}) where we first encountered the Clebsch-Gordan coefficients, but now involve the vector spherical harmonics. We will see, with some effort, that these integrals can be expressed entirely as combinations of Clebsch-Gordan coefficients. Note that ${\bf\Psi}_{l_3}^{m_3}\cdot{\bf\Psi}_{l_2}^{m_2}={\bf\Phi}_{l_3}^{m_3}\cdot{\bf\Phi}_{l_3}^{m_2}$, so the above two expressions exhaust the possibilities for what we call ``vector extensions" of the Clebsch-Gordan coefficients. 

First, we expand out the inner products of the vector spherical harmonics. For compactness, we are going to drop the argument of the spherical harmonics and their vector counterparts:
\begin{eqnarray}
\label{eq:innerProduct1}
{\bf\Phi}_{l_3}^{m_3}\cdot{\bf\Psi}_{l_2}^{m_2}&=&\frac{i}{2}\Bigg(\Lambda_+(l_2,m_2)\Lambda_-(l_3,m_3)\cos{\theta}Y_{l_3}^{m_3-1}Y_{l_2}^{m_2+1}+m_2\sin{\theta}\left[\Lambda_+(l_3,m_3)e^{-i\phi}Y_{l_3}^{m_3+1}Y_{l_2}^{m_2}-\Lambda_-(l_3,m_3)e^{i\phi}Y_{l_3}^{m_3-1}Y_{l_2}^{m_2}\right]-\nonumber\\
&&~~~~~~~\left\{2\leftrightarrow 3\right\}\Bigg);
\end{eqnarray}
\begin{eqnarray}
\label{eq:innerProduct2}
{\bf\Psi}_{l_3}^{m_3}\cdot{\bf\Psi}_{l_2}^{m_2}&=&-\frac{1}{4}\Bigg[\Lambda_+(l_3,m_3)\Lambda_+(l_2,m_2)\sin^2{\theta}e^{-2i\phi}Y_{l_3}^{m_3+1}Y_{l_2}^{m_2+1}+\Lambda_-(l_3,m_3)\Lambda_-(l_2,m_2)\sin^2{\theta}e^{2i\phi}Y_{l_3}^{m_3-1}Y_{l_2}^{m_2-1}-\nonumber\\
&&~~~~~~~~\left.\Bigg(\Lambda_+(l_3,m_3)\Lambda_-(l_2,m_2)(1+\cos^2{\theta})Y_{l_3}^{m_3+1}Y_{l_2}^{m_2-1}+\{2\leftrightarrow 3\}\Bigg)\right]-\nonumber\\
&&\frac{1}{2}\sin{\theta}\cos{\theta}\Bigg(m_3\Lambda_+(l_2,m_2)e^{-i\phi}Y_{l_3}^{m_3}Y_{l_2}^{m_2+1}+m_3\Lambda_-(l_2,m_2)e^{i\phi}Y_{l_3}^{m_3}Y_{l_2}^{m_2-1}+\{2\leftrightarrow 3\}\Bigg)+m_2m_3\sin^2{\theta}Y_{l_3}^{m_3}Y_{l_2}^{m_2}.\nonumber\\
\end{eqnarray}
We made use of the notation ``$\{2\leftrightarrow 3\}$" to indicate a copy of the preceding part of the parenthetically grouped expression with the indices $2$ and $3$ swapped; it only ever appears within a set of parentheses where it is clearly paired with a preceding expression. 

In order to mold these into a more desirable form, a form where all of the angular variation is encoded entirely in spherical harmonics, we make use of the recurrence relations in Eqs. (\ref{eq:recurrence1}) and (\ref{eq:recurrence2}) and the additional recurrence relation 
\begin{eqnarray}
(1-x^2)^{1/2}P_l^m(x)&=&-\frac{1}{(2l+1)}[P_{l+1}^{m+1}-P_{l-1}^{m+1}].
\end{eqnarray}
With these recurrence relations, we can derive the following helpful identities:
\begin{eqnarray}
\cos{\theta}Y_l^m&=&\sqrt{\frac{(l+m+1)(l-m+1)}{(2l+1)(2l+3)}}Y_{l+1}^m+\sqrt{\frac{(l+m)(l-m)}{(2l-1)(2l+1)}}Y_{l-1}^m;
\end{eqnarray}
\begin{eqnarray}
e^{-i\phi}\sin{\theta}Y_l^{m+1}&=&\sqrt{\frac{(l-m)(l-m+1)}{(2l+1)(2l+3)}}Y_{l+1}^m-\sqrt{\frac{(l+m)(l+m+1)}{(2l-1)(2l+1)}}Y_{l-1}^m;
\end{eqnarray}
\begin{eqnarray}
e^{i\phi}\sin{\theta}Y_l^{m-1}&=&-\sqrt{\frac{(l+m)(l+m+1)}{(2l+1)(2l+3)}}Y_{l+1}^m+\sqrt{\frac{(l-m)(l-m+1)}{(2l-1)(2l+1)}}Y_{l-1}^m.
\end{eqnarray}

Incorporating these into the inner products of vector spherical harmonics, we arrive at
\begin{eqnarray}
\Delta_{l_2,m_2;l_3,m_3}^{l_1,m_1}&=&\frac{i}{2}\Bigg(\sqrt{\frac{(2l_2+3)(2l_3+1)}{4\pi(2l_1+1)}}C_{l_2+1,0;l_3,0}^{l_1,0}\left[\Lambda_+(l_2,m_2)\Lambda_-(l_3,m_3)\sqrt{\frac{(l_2+m_2+2)(l_2-m_2)}{(2l_2+1)(2l_2+3)}}C_{l_2+1,m_2+1;l_3,m_3-1}^{l_1,m_1}-\right.\nonumber\\
&&~~~~~~~\left.m_3\left\{\Lambda_+(l_2,m_2)\sqrt{\frac{(l_2-m_2)(l_2-m_2+1)}{(2l_2+1)(2l_2+3)}}+\Lambda_-(l_2,m_2)\sqrt{\frac{(l_2+m_2)(l_2+m_2+1)}{(2l_2+1)(2l_2+3)}}\right\}C_{l_2+1,m_2;l_3,m_3}^{l_1,m_1}\right]+\nonumber\\
&&~~~\sqrt{\frac{(2l_2-1)(2l_3+1)}{4\pi(2l_1+1)}}C_{l_2-1,0;l_3,0}^{l_1,0}\left[\Lambda_+(l_2,m_2)\Lambda_-(l_3,m_3)\sqrt{\frac{(l_2+m_2+1)(l_2-m_2-1)}{(2l_2-1)(2l_2+1)}}C_{l_2-1,m_2+1;l_3,m_3-1}^{l_1,m_1}+\right.\nonumber\\
&&~~~~~~~\left.m_3\left\{\Lambda_+(l_2,m_2)\sqrt{\frac{(l_2+m_2)(l_2+m_2+1)}{(2l_2-1)(2l_2+1)}}+\Lambda_-(l_2,m_2)\sqrt{\frac{(l_2-m_2)(l_2-m_2+1)}{(2l_2-1)(2l_2+1)}}\right\}C_{l_2-1,m_2;l_3,m_3}^{l_1,m_1}\right]-\nonumber\\
&&~~~~~~~~~\{2\leftrightarrow 3\}\Bigg).
\end{eqnarray}
For $\Gamma_{l_2,m_2;l_3,m_3}^{l_1,m_1}$, it is helpful to break it into smaller pieces, i.e.,
\begin{eqnarray}
\Gamma_{l_2,m_2;l_3,m_3}^{l_1,m_1}&=&X_1+X_2+X_3+X_4+X_5,
\end{eqnarray}
where
\begin{eqnarray}
X_1&=&-\frac{1}{4}\int\Lambda_+(l_3,m_3)\Lambda_+(l_2,m_2)\sin^2{\theta}e^{-2i\phi}Y_{l_3}^{m_3+1}Y_{l_2}^{m_2+1}Y_{l_1}^{m_1*}d\Omega,\nonumber\\
&=&-\frac{\Lambda_+(l_3,m_3)\Lambda_+(l_2,m_2)}{4\sqrt{4\pi(2l_1+1)(2l_2+1)(2l_3+1)}}\bigg[\sqrt{(l_3-m_3)(l_3-m_3+1)(l_2-m_2)(l_2-m_2+1)}C_{l_2+1,0;l_3+1,0}^{l_1,0}C_{l_2+1,m_2;l_3+1,m_3}^{l_1,m_1}+\nonumber\\
&&~~~~~~~~~~~~~~~~~~~~~~~~~~~~~~~~~~~~~~~~~~~~~~~~~~~~~~~~\sqrt{(l_3+m_3)(l_3+m_3+1)(l_2+m_2)(l_2+m_2+1)}C_{l_2-1,0;l_3-1,0}^{l_1,0}C_{l_2-1,m_2;l_3-1,m_3}^{l_1,m_1}-\nonumber\\
&&~~~~~~~~~~~~~~~~~~~~~~~~~~~~~~~~~~~~~~~~~~~~~~~~~~~~~~~~\sqrt{(l_3-m_3)(l_3-m_3+1)(l_2+m_2)(l_2+m_2+1)}C_{l_2-1,0;l_3+1,0}^{l_1,0}C_{l_2-1,m_2;l_3+1,m_3}^{l_1,m_1}-\nonumber\\
&&~~~~~~~~~~~~~~~~~~~~~~~~~~~~~~~~~~~~~~~~~~~~~~~~~~~~~~~~\sqrt{(l_3+m_3)(l_3+m_3+1)(l_2-m_2)(l_2-m_2+1)}C_{l_2+1,0;l_3-1,0}^{l_1,0}C_{l_2+1,m_2;l_3-1,m_3}^{l_1,m_1}\bigg],
\end{eqnarray}
\begin{eqnarray}
X_2&=&-\frac{1}{4}\int\Lambda_-(l_3,m_3)\Lambda_-(l_2,m_2)\sin^2{\theta}e^{2i\phi}Y_{l_3}^{m_3-1}Y_{l_2}^{m_2-1}Y_{l_1}^{m_1*}d\Omega,\nonumber\\
&=&-\frac{\Lambda_-(l_3,m_3)\Lambda_-(l_2,m_2)}{4\sqrt{4\pi(2l_1+1)(2l_2+1)(2l_3+1)}}\bigg[\sqrt{(l_3+m_3)(l_3+m_3+1)(l_2+m_2)(l_2+m_2+1)}C_{l_2+1,0;l_3+1,0}^{l_1,0}C_{l_2+1,m_2;l_3+1,m_3}^{l_1,m_1}+\nonumber\\
&&~~~~~~~~~~~~~~~~~~~~~~~~~~~~~~~~~~~~~~~~~~~~~~~~~~~~~~~~\sqrt{(l_3-m_3)(l_3-m_3+1)(l_2-m_2)(l_2-m_2+1)}C_{l_2-1,0;l_3-1,0}^{l_1,0}C_{l_2-1,m_2;l_3-1,m_3}^{l_1,m_1}-\nonumber\\&&~~~~~~~~~~~~~~~~~~~~~~~~~~~~~~~~~~~~~~~~~~~~~~~~~~~~~~~~\sqrt{(l_3+m_3)(l_3+m_3+1)(l_2-m_2)(l_2-m_2+1)}C_{l_2-1,0;l_3+1,0}^{l_1,0}C_{l_2-1,m_2;l_3+1,m_3}^{l_1,m_1}-\nonumber\\
&&~~~~~~~~~~~~~~~~~~~~~~~~~~~~~~~~~~~~~~~~~~~~~~~~~~~~~~~~\sqrt{(l_3-m_3)(l_3-m_3+1)(l_2+m_2)(l_2+m_2+1)}C_{l_2+1,0;l_3-1,0}^{l_1,0}C_{l_2+1,m_2;l_3-1,m_3}^{l_1,m_1}\bigg],
\end{eqnarray}
\begin{eqnarray}
X_3&=&\frac{1}{4}\int\bigg(\Lambda_+(l_3,m_3)\Lambda_-(l_2,m_2)(1+\cos^2{\theta})Y_{l_3}^{m_3+1}Y_{l_2}^{m_2-1}+\{2\leftrightarrow 3\}\bigg)Y_{l_1}^{m_1*}d\Omega,\nonumber\\
&=&\frac{1}{4}\Bigg(\frac{\Lambda_+(l_3,m_3)\Lambda_-(l_2,m_2)}{\sqrt{4\pi(2l_1+1)(2l_2+1)(2l_3+1)}}\bigg[(2l_2+1)(2l_3+1)C_{l_2,0;l_3,0}^{l_1,0}C_{l_2,m_2-1;l_3,m_3+1}^{l_1,m_1}+\nonumber\\
&&~~~~\sqrt{(l_3+m_3+2)(l_3-m_3)(l_2+m_2)(l_2-m_2+2)}C_{l_2+1,0;l_3+1,0}^{l_1,0}C_{l_2+1,m_2-1;l_3+1,m_3+1}^{l_1,m_1}+\nonumber\\
&&~~~~\sqrt{(l_3+m_3+1)(l_3-m_3-1)(l_2+m_2-1)(l_2-m_2+1)}C_{l_2-1,0;l_3-1,0}^{l_1,0}C_{l_2-1,m_2-1;l_3-1,m_3+1}^{l_1,m_1}+\nonumber\\
&&~~~~\sqrt{(l_3+m_3+2)(l_3-m_3)(l_2+m_2-1)(l_2-m_2+1)}C_{l_2-1,0;l_3+1,0}^{l_1,0}C_{l_2-1,m_2-1;l_3+1,m_3+1}^{l_1,m_1}+\nonumber\\
&&~~~~\sqrt{(l_3+m_3+1)(l_3-m_3-1)(l_2+m_2)(l_2-m_2+2)}C_{l_2+1,0;l_3-1,0}^{l_1,0}C_{l_2+1,m_2-1;l_3-1;m_3+1}^{l_1,m_1}\bigg]+\{2\leftrightarrow3\}\Bigg),
\end{eqnarray}
\begin{eqnarray}
X_4&=&-\frac{1}{2}\int\sin{\theta}\cos{\theta}\bigg(m_3\Lambda_+(l_2,m_2)e^{-i\phi}Y_{l_3}^{m_3}Y_{l_2}^{m_2+1}+m_3\Lambda_-(l_2,m_2)e^{i\phi}Y_{l_3}^{m_3}Y_{l_2}^{m_2-1}+\{2\leftrightarrow 3\}\bigg)Y_{l_1}^{m_1*}d\Omega,\nonumber\\
&=&\frac{1}{2\sqrt{4\pi(2l_1+1)(2l_2+1)(2l_3+1)}}\Bigg(m_3\Bigg\{\bigg[\Lambda_+(l_2,m_2)\sqrt{(l_2+m_2)(l_2+m_2+1)}-\Lambda_-(l_2,m_2)\sqrt{(l_2-m_2)(l_2-m_2+1)}\bigg]\times\nonumber\\
&&~~~~~~~~~~~~~~~~~~~~~~~~~~~~~~~~~~~~~~~~~~~~~~~~~~~~~~~~~~~~~~~\sqrt{(l_3+m_3+1)(l_3-m_3+1)}C_{l_2-1,0;l_3+1,0}^{l_1,0}C_{l_2-1,m_2;l_3+1,m_3}^{l_1,m_1}+\nonumber\\
&&~~~~~~~~~~~~~~~~~~~~~~~~~~~~~~~~~~~~~~~~~~~~~~~~~~~~~~~~~\bigg[\Lambda_+(l_2,m_2)\sqrt{(l_2+m_2)(l_2+m_2+1)}-\Lambda_-(l_2,m_2)\sqrt{(l_2-m_2)(l_2-m_2+1)}\bigg]\times\nonumber\\
&&~~~~~~~~~~~~~~~~~~~~~~~~~~~~~~~~~~~~~~~~~~~~~~~~~~~~~~~~~\sqrt{(l_3+m_3)(l_3-m_3)}C_{l_2-1,0;l_3-1,0}^{l_1,0}C_{l_2-1,m_2;l_3-1,m_3}^{l_1,m_1}-\nonumber\\
&&~~~~~~~~~~~~~~~~~~~~~~~~~~~~~~~~~~~~~~~~~~~~~~~~~~~~~~~~~\bigg[\Lambda_+(l_2,m_2)\sqrt{(l_2-m_2)(l_2-m_2+1)}-\Lambda_-(l_2,m_2)\sqrt{(l_2+m_2)(l_2+m_2+1)}\bigg]\times\nonumber\\
&&~~~~~~~~~~~~~~~~~~~~~~~~~~~~~~~~~~~~~~~~~~~~~~~~~~~~~~~~~\sqrt{(l_3+m_3)(l_3-m_3)}C_{l_2+1,0;l_3-1,0}^{l_1,0}C_{l_2+1,m_2;l_3-1,m_3}^{l_1,m_1}-\nonumber\\
&&~~~~~~~~~~~~~~~~~~~~~~~~~~~~~~~~~~~~~~~~~~~~~~~~~~~~~~~~~\bigg[\Lambda_+(l_2,m_2)\sqrt{(l_2-m_2)(l_2-m_2+1)}-\Lambda_-(l_2,m_2)\sqrt{(l_2+m_2)(l_2+m_2+1)}\bigg]\times\nonumber\\
&&~~~~~~~~~~~~~~~~~~~~~~~~~~~~~~~~~~~~~~~~~~~~~~~~~~~~~~~~~\sqrt{(l_3+m_3+1)(l_3-m_3+1)}C_{l_2+1,0;l_3+1,0}^{l_1,0}C_{l_2+1,m_2;l_3+1,m_3;l_1,m_1}^{l_1,m_1}\bigg\}+\{2\leftrightarrow3\}\Bigg),
\end{eqnarray}
and
\begin{eqnarray}
X_5&=&m_2m_3\int \sin^2{\theta}Y_{l_3}^{m_3}Y_{l_2}^{m_2}Y_{l_1}^{m_1*}d\Omega,\nonumber\\
&=&\frac{m_2m_3}{\sqrt{4\pi(2l_1+1)(2l_2+1)(2l_3+1)}}\bigg[\sqrt{(l_3-m_3+1)(l_3-m_3+2)(l_2-m_2-1)(l_2-m_2)}C_{l_2-1,0;l_3+1,0}^{l_1,0}C_{l_2-1,m_2+1;l_3+1,m_3-1}^{l_1,m_1}+\nonumber\\
&&~~~~~~~~~~~~~~~~~~~~~~~~~~~~~~~~~~~~~~~~~~~~~\sqrt{(l_3+m_3-1)(l_3+m_3)(l_2+m_2+1)(l_2+m_2+2)}C_{l_2+1,0;l_3-1,0}^{l_1,0}C_{l_2+1,m_2+1;l_3-1,m_3-1}^{l_1,m_1}-\nonumber\\
&&~~~~~~~~~~~~~~~~~~~~~~~~~~~~~~~~~~~~~~~~~~~~~\sqrt{(l_3-m_3+1)(l_3-m_3+2)(l_2+m_2+1)(l_2+m_2+2)}C_{l_2+1,0;l_3+1,0}^{l_1,0}C_{l_2+1,m_2+1;l_3+1,m_3-1}^{l_1,m_1}-\nonumber\\
&&~~~~~~~~~~~~~~~~~~~~~~~~~~~~~~~~~~~~~~~~~~~~~\sqrt{(l_3+m_3-1)(l_3+m_3)(l_2-m_2-1)(l_2-m_2)}C_{l_2-1,0;l_3-1,0}^{l_1,0}C_{l_2-1,m_2+1;l_3-1,m_3-1}^{l_1,m_1}\bigg].
\end{eqnarray}
We note that just like standard Clebsch-Gordan coefficients---or, rather, because of this property of Clebch-Gordan coefficients--- $\Delta_{l_2,m_2;l_3,m_3}^{l_1,m_1}$ and $\Gamma_{l_2,m_2;l_3,m_3}^{l_1,m_1}$ vanish unless $m_1=m_2+m_3$.
\section{Appendix D: Temperature Perturbation from Memory-induced Deflections}
As we did with the BWM-induced redshift perturbation, we will carry out a multipolar decomposition of the BWM-induced deflection pattern. Since the deflection pattern is a vector field on the sphere, an appropriate basis for such a decomposition is formed by the vector spherical harmonics ${\bf \Phi}_l^m(\theta,\phi)$ and ${\bf \Psi}_l^m(\theta,\phi)$ which we describe in Eqs. (\ref{eq:Philm}) and (\ref{eq:Psilm}). In terms of these vector spherical harmonics, we can expand $\delta{\bf n}(t,\hat{\bf n})$ as
\begin{eqnarray}
\label{eq:vectorExpansion}
\delta{\bf n}(t,\hat{\bf n})&=&\sum_{l=1}^\infty\sum_{m=-l}^l[\rho_{lm}(t){\bf\Phi}_l^m(\hat{\bf n})+\sigma_{lm}(t){\bf\Psi}_l^m(\hat{\bf n})],
\end{eqnarray}
where
\begin{eqnarray}
\rho_{lm}(t)&=&\frac{1}{l(l+1)}\int\delta{\bf n}(t,\hat{\bf n})\cdot{\bf\Phi}_l^m(\hat{\bf n})d\Omega,~{\rm and}\\
\sigma_{lm}(t)&=&\frac{1}{l(l+1)}\int\delta{\bf n}(t,\hat{\bf n})\cdot{\bf\Psi}_l^m(\hat{\bf n})d\Omega.
\end{eqnarray}
We note that the summations in Eq. (\ref{eq:vectorExpansion}) begins at $l=1$ because ${\bf\Phi}_0^0(\hat{\bf n})={\bf\Psi}_0^0(\hat{\bf n})=0$. From our expressions for ${\bf V}_\oplus$ and ${\bf V}_\bigstar$ in Eqs. (\ref{eq:earthterm}) and (\ref{eq:starterm}) and the form of the vector spherical harmonics given in Eqs. (\ref{eq:Philm}) and (\ref{eq:Psilm}), it is clear that $\rho_{lm}$ and $\sigma_{lm}$ will vanish from the azimuthal integration unless $m=\pm2$. 

Here we compute $\rho_{l,+2}$ and $\sigma_{l,+2}$. It is helpful to break these calculations into manageable chunks. To that end, we write
\begin{eqnarray}
\rho_{l,+2}(t)&=&\frac{h_M}{l(l+1)}[{\cal R}_1+{\cal R}_2(t)+{\cal R}_3(t)],~~{\rm and}\\
\sigma_{l,+2}(t)&=&\frac{h_M}{l(l+1)}[{\cal S}_1+{\cal S}_2(t)+{\cal S}_3(t)],
\end{eqnarray}
where
\begin{eqnarray}
{\cal R}_1&=&\int {\bf V}_\oplus(\hat{\bf n})\cdot{\bf\Phi}_l^{+2}(\hat{\bf n})~d\Omega,\\
{\cal R}_2(t)&=&-\int\frac{\beta\Theta(\theta_t-\theta)}{(1+\cos{\theta})}{\bf V}_\bigstar(\hat{\bf n})\cdot{\bf\Phi}_l^{+2}(\hat{\bf n})~d\Omega,\\
{\cal R}_3(t)&=&-\int\Theta(\theta-\theta_t){\bf V}_\bigstar(\hat{\bf n})\cdot{\bf\Phi}_l^{+2}(\hat{\bf n})~d\Omega,\\
{\cal S}_1&=&\int {\bf V}_\oplus(\hat{\bf n})\cdot{\bf\Psi}_l^{+2}(\hat{\bf n})~d\Omega,\\
{\cal S}_2(t)&=&-\int\frac{\beta\Theta(\theta_t-\theta)}{(1+\cos{\theta})}{\bf V}_\bigstar(\hat{\bf n})\cdot{\bf\Psi}_l^{+2}(\hat{\bf n})~d\Omega,~~{\rm and}\\
{\cal S}_3(t)&=&-\int\Theta(\theta-\theta_t){\bf V}_\bigstar(\hat{\bf n})\cdot{\bf\Psi}_l^{+2}(\hat{\bf n})~d\Omega.
\end{eqnarray}
With the tools developed in these appendixes, these integrals can be worked out explicitly:
\begin{eqnarray}
{\cal R}_1&=&\frac{1}{8}\sqrt{\pi(2l+1)}\left\{4\gamma(l,+2)\left[{\cal I}_{G,l}(-1,1)-{\cal I}_{I,l}(-1,1)\right]-\right.\nonumber\\
&&~~~~~~~~~~~~~~\Lambda_+(l,+2)\gamma(l,+3)[{\cal I}_{L,l}(-1,1)-{\cal I}_{K,l}(-1,1)]-\nonumber\\
&&~~~~~~~~~~~~~\left.\Lambda_-(l,+2)\gamma(l,+1)[{\cal I}_{D,l}(-1,1)+{\cal I}_{E,l}(-1,1)]\right\};
\end{eqnarray}
\begin{eqnarray}
{\cal R}_2(t)&=&-\frac{\beta}{8}\sqrt{\pi(2l+1)}\left\{4\gamma(l,+2)[{\cal I}_{G,l}(\beta-1,1)-{\cal I}_{I,l}(\beta-1,1)]-\right.\nonumber\\
&&~~~~~~~~~~~~~\Lambda_+(l,+2)\gamma(l,+3)[{\cal I}_{L,l}(\beta-1,1)-2{\cal I}_{N,l}(\beta-1,1)]-\nonumber\\
&&~~~~~~~~~~~~~\left.\Lambda_-(l,+2)\gamma(l,+1)[{\cal I}_{E,l}(\beta-1,1)+2{\cal I}_{A,l}(\beta-1,1)-2{\cal I}_{B,l}(\beta-1,1)]\right\};
\end{eqnarray}
\begin{eqnarray}
{\cal R}_3(t)&=&-\frac{1}{8}\sqrt{\pi(2l+1)}\left\{4\gamma(l,+2)[{\cal I}_{G,l}(-1,\beta-1)+{\cal I}_{H,l}(-1,\beta-1)-{\cal I}_{I,l}(-1,\beta-1)-{\cal I}_{J,l}(-1,\beta-1)]\right.-\nonumber\\
&&~~~~~~~~~~~~~~~\Lambda_+(l,+2)\gamma(l,+3)[{\cal I}_{L,l}(-1,\beta-1)+{\cal I}_{M,l}(-1,\beta-1)-2{\cal I}_{K,l}(-1,\beta-1)]-\nonumber\\
&&~~~~~~~~~~~~~~~\left.\Lambda_-(l,+2)\gamma(l,+1)[{\cal I}_{E,l}(-1,\beta-1)+{\cal I}_{F,l}(-1,\beta-1)+2{\cal I}_{D,l}(-1,\beta-1)]\right\};
\end{eqnarray}
\begin{eqnarray}
{\cal S}_1&=&-i~{\cal R}_1
\end{eqnarray}
\begin{eqnarray}
{\cal S}_2(t)&=&\frac{i\beta}{8}\sqrt{\pi(2l+1)}\left\{8\gamma(l,+2)[{\cal I}_{G,l}(\beta-1,1)-{\cal I}_{H,l}(\beta-1,1)]\right.+\nonumber\\
&&~~~~~~~~~~~~~\Lambda_+(l,+2)\gamma(l,+3)[{\cal I}_{N,l}(\beta-1,1)-{\cal I}_{O,l}(\beta-1,1)]-\nonumber\\
&&~~~~~~~~~~~~\left.\Lambda_-(l,+2)\gamma(l,+1)[{\cal I}_{A,l}(\beta-1,1)+2{\cal I}_{B,l}(\beta-1,1)-3{\cal I}_{C,l}(\beta-1,1)]\right\};
\end{eqnarray}
\begin{eqnarray}
{\cal S}_3(t)&=&\frac{i}{8}\sqrt{\pi(2l+1)}\left\{8\gamma(l,+2)[{\cal I}_{G,l}(-1,\beta-1)-{\cal I}_{I,l}(-1,\beta-1)]\right.+\nonumber\\
&&~~~~~~~~~~~~~~~~~~~~~~~\Lambda_+(l,+2)\gamma(l,+3)[{\cal I}_{K,l}(-1,\beta-1)-{\cal I}_{L,l}(-1,\beta-1)]-\nonumber\\
&&~~~~~~~~~~~~~~~~~~~~~~\Lambda_-(l,+2)\gamma(l,+1)[{\cal I}_{D,l}(-1,\beta-1)+3{\cal I}_{E,l}(-1,\beta-1)].
\end{eqnarray}

We see that $\rho_{l,+2}(t)$ is entirely real and that $\sigma_{l,+2}(t)$ is entirely imaginary. It can be shown that $\rho_{l,-2}=-\rho_{l,+2}$ and $\sigma_{l,-2}=\sigma_{l,+2}$, allowing us to write Eq. (\ref{eq:vectorExpansion}) as
\begin{eqnarray}
\label{eq:deltansimple}
\delta{\bf n}(t,\hat{\bf n})&=&\sum_{l=2}^\infty\rho_{l,+2}(t)[{\bf\Phi}_l^{+2}(\hat{\bf n})-{\bf\Phi}_l^{-2}(\hat{\bf n})]+\sigma_{l,+2}(t)[{\bf\Psi}_l^{+2}(\hat{\bf n})+{\bf\Psi}_l^{-2}(\hat{\bf n})].
\end{eqnarray}
It can further be shown that ${\bf\Phi}_l^{-2}=-{\bf\Phi}_l^{+2*}$ and ${\bf\Psi}_l^{-2}=-{\bf\Psi}_l^{+2*}$. With this, we can again rewrite Eq. (\ref{eq:vectorExpansion}) as
\begin{eqnarray}
\delta{\bf n}(t,\hat{\bf n})&=&\sum_{l=2}^\infty\rho_{l,+2}(t)[{\bf\Phi}_l^{+2}(\hat{\bf n})+{\bf\Phi}_l^{+2*}(\hat{\bf n})]+\sigma_{l,+2}(t)[{\bf\Psi}_l^{+2}(\hat{\bf n})-{\bf\Psi}_l^{+2*}(\hat{\bf n})].
\end{eqnarray} 
This form makes it clear that $\delta{\bf n}(t,\hat{\bf n})$ is entirely real, an important consistency check.

Using the form of $\delta{\bf n}(t,\hat{\bf n})$ in Eq. (\ref{eq:deltansimple}), we can now write
\begin{eqnarray}
\label{eq:perpCoeff}
\delta a^\perp_{l_1m_1}(t)=-i\sum_{l_2=1}^\infty\sum_{m_2=-l_2}^{l_2}a_{l_2m_2}^0\sum_{l_3=2}^\infty\int\left\{\rho_{l_3,+2}(t)[{\bf\Phi}_{l_3}^{+2}(\hat{\bf n})-{\bf\Phi}_{l_3}^{-2}(\hat{\bf n})]+\sigma_{l_3,+2}(t)[{\bf\Psi}_{l_3}^{+2}(\hat{\bf n})+{\bf\Psi}_{l_3}^{-2}(\hat{\bf n})]\right\}\cdot{\bf\Psi}_{l_2}^{m_2}(\hat{\bf n})Y_{l_1}^{m_1*}(\hat{\bf n})d\Omega.\nonumber\\
\end{eqnarray}
Here we make use of our earlier calculation of what we referred to as vector extensions of the Clebsch-Gordan coefficients, $\Delta_{l_2,m_2;l_3,m_3}^{l_1,m_1}$ and $\Gamma_{l_2,m_2;l_3,m_3}^{l_1,m_1}$. The fact that these quantities vanish unless $m_1=m_2+m_3$ allows us to eliminate the summation over $m_2$ in Eq. (\ref{eq:perpCoeff}):
\begin{eqnarray}
\label{eq:deflectionPert}
\delta a^\perp_{l_1,m_1}&=&-i\sum_{l_2=1}^\infty\sum_{l_3=2}^\infty\bigg\{\rho_{l_3,+2}(t)\left[a_{l_2,m_1-2}^0\Delta_{l_2,m_1-2;l_3,+2}^{l_1,m_1}-a_{l_2,m_1+2}^0\Delta_{l_2,m_1+2;l_3,-2}^{l_1,m_1}\right]+\nonumber\\&&~~~~~~~~~~~~~~~~~~~~\sigma_{l_3,+2}(t)\left[a_{l_2,m_1-2}^0\Gamma_{l_2,m_1-2;l_3,+2}^{l_1,m_1}+a_{l_2,m_1+2}^0\Gamma_{l_2,m_1+2;l_3,-2}^{l_1,m_1}\right]\bigg\}.
\end{eqnarray}
This result, by construction, closely mirrors the form of Eq. (\ref{eq:redshiftPert}). Everything in it has been computed in these appendixes. One meaningful departure from Eq. (\ref{eq:redshiftPert}) in this expression is that the summation over $l_2$ begins at $l_2=1$ rather than $l_2=0$. This means the redshift perturbation from a BWM is capable of transferring power from the monopole of the CMB into higher degree modes but the deflection perturbation is not. This fact is demonstrated in Figs. \ref{fig:deltaAlmParallel}--\ref{fig:deflectionPatterns} and discussed in Sec. V. 
\vspace{.25cm}
\begin{acknowledgments}
D. R. M. is a member of the NANOGrav Physics Frontiers Center and was supported through the National Science Foundation NANOGrav Physics Frontiers Center Grant No. 1430284 during the initial stages of this project. D. R. M. is now supported through a Cottrell Fellowship with Research Corporation for Science Advancement. The Cottrell Fellowship Initiative is partially funded by the National Science Foundation under Grant No. CHE-2039044.
\end{acknowledgments}

\bibliographystyle{apsrev4-1}
\bibliography{spectrum.bib}
\end{document}